\documentclass{article}
\usepackage{amssymb}
\usepackage{epsfig}
\usepackage{amsmath}

\input{tcilatex}

\begin{document}

\title{\textbf{Frame-dragging and bending of Light 
in Kerr and Kerr-(anti) de Sitter spacetimes.}}
\author{G. V. Kraniotis \footnote{kraniotis@physics.tamu.edu} \\
George P. and Cynthia W. Mitchell Institute for Fundamental Physics,\\
Texas A$\&$M University\\
College Station TX,77843 USA \\
\footnote{
MIFP-05-16, July 2005}, \\
}
\maketitle

\begin{abstract}
The equations of general relativity in the form of timelike and null geodesics that describe motion of test particles and 
photons in Kerr spacetime are solved exactly including the contribution from 
the cosmological constant. We then perform a systematic application of the 
exact solutions obtained to the following cases.
The exact solutions derived for null, spherical, polar and non-polar  orbits are applied for the calculation of frame dragging (Lense-Thirring effect) for the orbit of a photon around the galactic centre, assuming that 
the latter is a Kerr black hole for various values of the Kerr parameter including those supported by recent observations.
Unbound null polar orbits are investigated, and an analytical expression 
for the deviation  angle of a polar photon orbit from the gravitational Kerr 
field is derived.
In addition, we present the exact solution for timelike and null equatorial 
orbits. In the former case, we derive an analytical expression 
for the precession of the point of closest approach 
(perihelion, periastron) for the orbit of a test particle around 
a rotating mass whose surrounding curved spacetime geometry is 
described by the Kerr field.
In the latter case, we calculate an exact 
expression for the deflection angle for a light ray in the gravitational 
field of a rotating mass (the Kerr field). We apply this calculation for 
the bending of light  from the gravitational field of the galactic centre, 
for various values of the Kerr parameter,  and the impact factor.

\end{abstract}

\bigskip

\bigskip \newpage

\section{ Introduction}

\subsection{\protect\bigskip Motivation}

Most of the celestial bodies deviate very little from spherical
symmetry, and the Schwarzschild spacetime is an appropriate
approximation for their gravitational field \cite{Karl}. However, 
for some astrophysical bodies the rotation of the mass distribution 
cannot be neglected. A more general spacetime solution of the
gravitational field equations should take this property into account.
In this respect, the Kerr solution \cite{KERR} represents, the curved 
spacetime 
geometry surrounding a rotating mass \cite{OHANIAN}. Moreover, the above 
solution 
is also important  for probing the strong 
field regime of general relativity \cite{REES}.
This is significant, since general relativity has triumphed in large-scale 
cosmology \cite{PERLMUTTER, BAHCALL, DEBERNAR,
GVKSBW}, and in predicting solar system effects on planetary orbits 
like the perihelion 
precession of Mercury with a very high precision \cite{Albert, 
Mercury} \footnote{See also about the BepiColombo science mission on Mercury http://sci.esa.int/home/bepicolombo/}.

As was discussed in \cite{KraniotisKerr}, the investigation of 
spacetime structures near strong gravitational sources, like neutron stars or 
candidate black hole (BH) systems is of paramount importance for testing the 
predictions of the theory in the strong field regime. The study of geodesics 
are crucial in this respect, in providing information of the 
structure of spacetime in the strong field limit.

The study of the geodesics from the Kerr metric are additionally motivated by
recent observational evidence of stellar orbits around the galactic centre, 
which indicates that the spacetime surrounding the $\rm{Sgr\; A^{*}}$ radio source,
which is believed to be a supermassive black hole of 3.6 million solar
masses, is described by the Kerr solution rather than the Schwarzschild 
solution, with the Kerr parameter \cite{GENZEL} 
\begin{equation}
\frac{J}{G M_{{\rm BH}}/c}=0.52\; (\pm 0.1,\pm 0.08,\pm 0.08)
\label{Galaxy}
\end{equation}
where the reported high-resolution infrared 
observations of $\rm {Sgr\; A^{*}}$
revealed `quiescent' emission and several flares.
This is half the maximum value for a Kerr black hole \cite{SPIN}. 
In the above equation $J$ \footnote{$J=c a$ where $a$ is the Kerr parameter. The interpretation of $ca$ as the angular momentum per unit mass was first given by Boyer and Price \cite{BOPR}. In fact, by comparing with the Lense-Thirring 
calculations \cite{LTPr} they determined the Kerr parameter to be: $a=-
\frac{2 \Omega l^2}{5 c}$, where $\Omega$ and $l$ denote the angular 
velocity and radius of the rotating sphere.}  
denotes the angular momentum of the 
black hole (The error 
estimates here the uncertainties in the period, black hole mass ($M_{{\rm BH}}$) 
and distance to the galactic centre, respectively; $G$ is the gravitational 
constant and $c$ the velocity of light.)

Taking into account the cosmological constant $\Lambda$ contribution,
the generalization of the Kerr solution is described by the Kerr -de Sitter
metric element which in Boyer-Lindquist (BL) coordinates \footnote{These 
coordinates have the advantage that reduce to the Schwarzschild solution 
with a cosmological constant in the limit $a\rightarrow 0$, see \cite{Boyer}.}
is given by \cite{Zdenek,CARTER}:

\begin{eqnarray}
ds^2&=&\frac{\Delta_r}{\Xi^2 \rho^2}\left(c dt-a\sin^2
\theta d\phi\right)^2
-\frac{\rho^2}{\Delta_r}dr^2-\frac{\rho^2}{\Delta_{\theta}}d\theta^2\nonumber
\\
&-&\frac{\Delta_{\theta}\sin^2
\theta}{\Xi^2 \rho^2}\left(a c dt-(r^2+a^2)d\phi\right)^2
\end{eqnarray}
where 
\begin{eqnarray}
\Delta_r &:=&(1-\frac{\Lambda}{3}r^2)(r^2+a^2)-\frac{2 G M r}{c^2} \nonumber \\
\Delta_{\theta}&:=&1+\frac{a^2 \Lambda}{3}\cos^2 \theta \nonumber \\
\Xi&:=&1+\frac{a^2 \Lambda}{3},\;\;\rho^2:=r^2+a^2  \cos^2 \theta \nonumber \\
 \end{eqnarray}

In a recent paper \cite{KraniotisKerr}, we derived the timelike geodesic equations in Kerr 
spacetime with a cosmological constant by solving the Hamilton-Jacobi 
partial differential equation by separation of variables. 
Subsequently, we solved exactly the corresponding differential equations 
for an interesting class of possible types of motion for a test particle in 
Kerr and Kerr-(anti) de Sitter spacetimes. 
The exact solution of non-spherical geodesics was obtained by using the 
transformation theory of elliptic functions.

The exact solutions of the timelike geodesic equations  obtained 
in \cite{KraniotisKerr} were applied to the following 
situations:

{\em Frame dragging from rotating gravitational mass}. An essential property of the geodesics in Schwarzschild
spacetime is that although the orbit precesses relativistically it remains in the same plane; the Kerr rotation adds longitudinal dragging 
to this precession.  For instance, in the {\em spherical polar 
orbits} we discussed in \cite{KraniotisKerr}, (where the particle 
traverses all latitudes, passes through the symmetry axis $z$, infinitely many times) the angle of longitude increases  after a complete oscillation in latitude.  This phenomenon, is in accordance with {\em Mach's principle}.

More specifically, in \cite{KraniotisKerr} we calculated 
 the dragging of inertial frames in the following situations.
(a) Dragging of a satellite's spherical polar orbit in the 
gravitational field of Earth 
assuming Kerr geometry, using as radii, the semi-major axis of the 
polar orbit of the GP-B mission \footnote{http://einstein.stanford. See also 
\cite{Everitt}.} 
launched in April 2004. 
(b) Dragging of a stellar, spherical polar orbit, in the gravitational field 
of a rotating galactic black hole.

It is the purpose of this paper to extend the analysis and applications 
of the exact solutions obtained in \cite{KraniotisKerr} to an interesting 
class of possible types of motion for a test 
particle in Kerr and Kerr-(anti) de Sitter spacetimes, as well as to 
derive the exact solutions of null geodesics in the same spacetimes 
and explore their physical implications.
In the latter case, we apply the 
exact solutions obtained, to the following situations:

(a) Dragging of a photon's spherical polar and non-polar orbit in the 
gravitational field of a rotating galactic centre black hole.

(b) The deflection angle of a light ray from the gravitational field of a 
rotating  black hole, for various values of the Kerr parameter and the impact factor.

The material of this paper is organized as follows.
In section \ref{Integrability} we review the derivation of the 
relevant geodesic equations. In section \ref{LauricellaHyp} we 
discuss the definition of the Lauricella's hypergeometric function 
of many variables, as well as the integral representations that it admits, 
which are important in our exact treatment of geodesic equations that 
describe motion of a test particle and photon 
in Kerr-(anti) de Sitter spacetime.
In sections \ref{PSPLAMBDA},\ref{SPNULL} and \ref{SNP}, we solve exactly spherical polar or non-polar null geodesics 
with and without the cosmological constant.
In the case of spherical polar photonic geodesics and for 
 a vanishing cosmological 
constant the exact solution for the orbit  is given by the Weierstra$\ss$ 
elliptic function. The exact expression for frame dragging is proportional 
to the real half-period of the Weierstra$\ss$ modular Jacobi form.
In the case of non-polar spherical photonic orbits, the exact expression 
for the Lense-Thirring precession of the photon is given in terms of 
a hypergeometric function of one variable and Appell's first generalized 
hypergeometric function of two variables $F_1$ \cite{APPELL}. Assuming a vanishing cosmological constant and that the galactic centre is a supermassive rotating black hole, we apply the exact solutions obtained for the determination 
of Lense-Thirring precession for a photon in spherical polar and non-polar 
orbits around the galactic centre. The corresponding exact expressions in 
the presence of the cosmological constant are also derived and discussed in 
sections \ref{PSPLAMBDA} and \ref{SNPLAMBDA}.


In section \ref{NSP} we perform a precise calculation for the deflection angle 
of a  photon's $non-spherical$ {\em polar} 
orbit from the gravitational Kerr field. In this novel case, the exact expressions obtained were written in terms of Lauricella's hypergeometric function 
$F_D$.

Timelike spherical polar  (with $\Lambda\not =0$) and non-polar (with 
$\Lambda=0$) orbits 
are treated in sections 
\ref{TPOLARFRAME}, \ref{LNZSPHE} respectively.

In sections \ref{TIMEEQUATOR} and 
\ref{PHOSEQUATOR} we study the exact solution of non-spherical timelike and 
null equatorial orbits respectively.
The amount of relativistic precession for a test particle in a 
timelike orbit, confined to the equatorial plane, in the presence of rotation 
of the central mass is given 
in terms of Appell's first hypergeometric function of two variables 
$F_1$. On the other hand, the exact expression for the 
deflection angle of a photonic orbit from the Kerr gravitational field 
surrounding a rotating 
central mass is given in terms of Appell's $F_1$ hypergeometric function  and 
Lauricella's fourth hypergeometric function $F_D$ of three variables 
\cite{LAURICELLA}.
We use section \ref{SYMBE} for our conclusions. In the appendices, we 
collect some of our formal calculations, as well as some useful properties 
of Appell's hypergeometric function and definitions of genus-2 theta functions.

\section{Separability of Hamilton-Jacobi's differential equation in 
Kerr-(anti) de Sitter metric and derivation of geodesics.}
\label{Integrability}

In the presence of the cosmological constant it was proved in 
\cite{KraniotisKerr} the important 
result that the Hamilton-Jacobi differential equation 
can be solved by separation of variables. Thus in this case, the 
 characteristic function separates and takes the form \cite{KraniotisKerr}
\begin{eqnarray}
  W&=&-Ect +L\phi+\int\frac{\sqrt{\left[Q+(L-a E)^2 \Xi^2-
\mu^2 a^2 \cos^2\theta\right]\Delta_{\theta}-\frac{\Xi^2(a E\sin^2\theta-L)^2}{\sin^2\theta}}}{\Delta_{\theta}}d\theta \nonumber \\
&+&\int\frac{\sqrt{\Xi^2\left[(r^2+a^2)E-aL\right]^2-\Delta_r(\mu^2r^2+Q+
\Xi^2(L-a E)^2)}}{\Delta_r}dr \nonumber 
\end{eqnarray}

By differentiating now with respect to constants of integration, $Q,L,E,\mu$, 
we obtain the following set of geodesic differential equations
\begin{eqnarray}
\int \frac{dr}{\sqrt{R^{\prime}}}=\int \frac{d\theta}{\sqrt{\Theta^{\prime}}}
\nonumber \\
\rho^2 \frac{d\phi}{d\lambda}=-\frac{\Xi^2}{\Delta_{\theta}\sin^2\theta}
\left(aE\sin^2\theta-L\right)+\frac{a\Xi^2}{\Delta_r}\left[(r^2+a^2)E-aL\right]
\nonumber \\
c\rho^2 \frac{dt}{d\lambda}=\frac{\Xi^2(r^2+a^2)\left[(r^2+a^2)E-aL\right]}{\Delta_r}-\frac{a\Xi^2 (aE \sin^2\theta-L)}{\Delta_{\theta}} \nonumber \\
\rho^2\frac{dr}{d\lambda}=\pm \sqrt{R^{\prime}} \nonumber \\
\rho^2\frac{d\theta}{d\lambda}=\pm \sqrt{\Theta^{\prime}} 
\label{LambdaGeo}
\end{eqnarray}
where 
\begin{eqnarray}
R^{\prime}&:=&\Xi^2 \left[(r^2+a^2)E-aL\right]^2-\Delta_r\left(\mu^2r^2+Q+
\Xi^2(L-aE)^2\right) \nonumber \\
\Theta^{\prime}&:=&\left[Q+(L-aE)^2 \Xi^2-\mu^2a^2 \cos^2\theta\right]\Delta_{\theta}-\Xi^2\frac{(aE \sin^2\theta-L)^2}{\sin^2\theta}
\label{LARTH}
\end{eqnarray}

The first line of Eq.(\ref{LambdaGeo}) is a differential equation that relates 
a {\em hyperelliptic} Abelian integral to an elliptic integral  which is the generalisation 
of the theory of transformation of elliptic functions discussed in 
\cite{KraniotisKerr}, in the case of non-zero cosmological constant. The mathematical treatment 
of such a relationship was first discussed by Abel in \cite{NHA2}.

Assuming a zero cosmological constant, as was shown by Carter, 
one gets
\begin{equation}
W=-Ect+\int\frac{\sqrt{R}}{\Delta}dr+\int \sqrt{\Theta}d\theta+L \phi
\label{SepC}
\end{equation}
where
\begin{equation}
\Theta:=Q-\left[a^2(\mu^2-E^2)+\frac{L^2}{\sin^2\theta}\right]\cos^2\theta
\label{poliki}
\end{equation}
and 
\begin{equation}
R:=\left[ (r^2+a^2)E-a L\right]^2-\Delta \left[ \mu^2 r^2+ (L-a E)^2+ Q\right]
\end{equation}
with $\Delta:=r^2+a^2-\frac{2 G M r}{c^2}$. Also $E,L$ are constants of integration associated with the isometries of the Kerr metric. Carter's constant of 
integration is denoted by $Q$.  
Differentiation of (\ref{SepC}), with respect to the integration constants $E,
L,Q,\mu$ leads to the  following set of first-order equations of motion
\cite{carter2}:
\begin{eqnarray}
 \rho^2 \frac{c dt}{d\lambda}&=&\frac{r^2+a^2}{\Delta}P-a\left(a\; E\; \sin^2 \theta-L\right) \nonumber \\
 \rho^2 \frac{dr}{d\lambda}&=&\pm \sqrt{R} \nonumber \\
 \rho^2 \frac{d\theta}{d\lambda}&=&\pm \sqrt{\Theta} \nonumber \\ 
 \rho^2 \frac{d\phi}{d\lambda}&=&\frac{a}{\Delta}P-aE+\frac{L}{\sin^2 \theta} 
\label{kerrgeo}
\end{eqnarray}
where 
\begin{equation}
P:=E (r^2+a^2)-a L
\label{GenikiP}
\end{equation}

Null-geodesics are derived by setting $\mu=0$.

\subsection{Lauricella's multivariable hypergeometric functions}
\label{LauricellaHyp}
Giuseppe Lauricella, building on the work of Appell 
who had developed hypergeometric functions of two variables, investigated 
in a systematic way, multiple 
hypergeometric functions at the end of the nineteenth century \cite{LAURICELLA}. 
He defined, 
four functions which are named after him and have both multiple series 
and integral representations. In particular, the fourth of these functions,
denoted by $F_D$, admits integral representations of importance in our 
exact treatment of geodesic equations, which describe motion of a test 
particle in Kerr-(anti) de Sitter spacetime.

The fourth Lauricella function of $m$-variables is given by
\begin{eqnarray}
F_D(\alpha,{\bf \beta},\gamma;{\bf z})&=&
\sum_{n_1,n_2,\cdots,n_m=0}^{\infty}\frac{(\alpha)_{n_1+\cdots+n_m}
(\beta_1)_{n_1}\cdots(\beta_m)_{n_m}}{(\gamma)_{n_1+\cdots+n_m}(1)_{n_1}\cdots(1)_{n_m}}z_1^{n_1}\cdots z_m^{n_m} \nonumber \\
&=&\sum_{n_1,n_2,\cdots,n_m=0}^{\infty}
\frac{\sum(\alpha,n_1+\cdots n_m)(\beta_1,n_1)\cdots(\beta_m,n_m)z_1^{n_1}\cdots
z_m^{n_m}}{(\gamma,n_1+\cdots +n_m)n_1!\cdots n_m!} \nonumber
\end{eqnarray}
where 
\begin{eqnarray}
{\bf z}&=&(z_1,\cdots,z_m) \nonumber \\
{\bf {\beta}}&=&(\beta_1,\cdots,\beta_m)
\end{eqnarray}
The Pochhammer symbol $(\alpha)_m=(\alpha,m)$ is defined by
$$(\alpha)_m=\frac{\Gamma(\alpha+m)}{\Gamma(\alpha)}=\Bigg \{ \begin{array}{ll}
1,& {\rm if}\;\; m=0, \\
\alpha(\alpha+1)\cdots (\alpha+m-1), & {\rm if}\;\; m=1,2,3,\cdots
\end{array}
$$

The series admits the following integral representation
\begin{equation}
F_D(\alpha,{\bf{\beta}},\gamma;{\bf z})=
\frac{\Gamma(\gamma)}{\Gamma(\alpha)\Gamma(\gamma-\alpha)}
\int_0^1 t^{\alpha-1}(1-t)^{\gamma-\alpha-1}(1-z_1 t)^{-\beta_1}\cdots
(1-z_m t)^{-\beta_m } dt
\end{equation}
which is valid for ${\rm Re}(\alpha)>0, {\rm Re}(\gamma-\alpha)>0$. It converges absolutely inside the $m$-dimensional cuboid:
\begin{equation}
|z_j|<1,\;\;\;\;\;(j=1,\cdots,m)
\end{equation}

\section{Spherical polar null geodesics}

Depending on whether or not the coordinate radius $r$ is constant along a 
given geodesic, the corresponding particle orbit is characterized 
as spherical or non-spherical respectively. In this section, we will 
concentrate on spherical polar photon orbits with a vanishing cosmological 
constant. We should mention at this point the extreme black hole solutions 
$a=1$ of spherical non-polar photon geodesics obtained in \cite{Teo} in 
terms of formal integrals.

The exact solution of the corresponding timelike orbits and their 
physical applications have been derived and investigated
in \cite{KraniotisKerr}.

Assuming a zero cosmological constant, $r=r_f$, where 
$r_f$ is a constant value setting $\mu=0$ and using the last two 
equations of (\ref{kerrgeo}) we  
obtain:

\begin{equation}
\frac{d\phi}{d\theta}=\frac{\frac{a P}{\Delta}-a E+L/\sin^2\theta}{\sqrt{\Theta}}
\label{PolargeodN}
\end{equation} 
where $\Theta$ now is given by
\begin{equation}
\Theta=Q-[-a^2 E^2 +\frac{L^2}{\sin^2 \theta}]\cos^2 \theta
\end{equation}

It is convenient to introduce the parameters
\begin{equation}
\Phi:=L/E,\;\;\;{\cal Q}:=Q/E^2
\label{PAPHI}
\end{equation}

Now by defining $z:=\cos^2\theta$, the previous equation can be written as 
follows,
\begin{equation}
d\phi=-\frac{1}{2}\frac{dz}{\sqrt{z^3\alpha-z^2(\alpha+\beta)+Q z}}\times \left\{
\frac{a P}{\Delta}-a + \frac{\Phi}{1-z}\right\}
\label{Kugel}
\end{equation}
where 
\begin{equation}
\alpha:=-a^2 ,\;\;\beta:={\cal Q}+\Phi^2
\end{equation}
It has been shown \cite{StoTsou} that a necessary condition  for an orbit to be $polar$ (meaning to intersect the symmetry axis of the 
Kerr gravitational field)  is the vanishing of the parameter $L$, i.e . $L=0$.
Assuming $\Phi=0$, in equation (\ref{Kugel}), we can transform it into the Weierstra$\ss$ form of an elliptic curve by the 
following substitution
\begin{equation}
z:=-\frac{\xi+\frac{\alpha+\beta}{12}}{-\alpha/4}
\label{subwei}
\end{equation}
Thus, we obtain the integral equation
\begin{equation}
\int d\phi=\int -\frac{1}{2}\frac{d\xi}{\sqrt{4 \xi^3-g_2 \xi-g_3}}\times
\left\{
\frac{a P^{\prime}}{\Delta}-a \right\}
\end{equation}
and this orbit integral can be $inverted$ by the Weierstra$\ss$ modular 
Jacobi form \footnote{For more information on the properties 
of the Weierstra$\ss$ function, the reader is referred to the monographs 
\cite{DON,Silverman}, and the appendix of \cite{GVKSBW}.}
\begin{equation}
\xi=\wp\left(\phi/A\right)
\end{equation} 
where $A:=-\frac{1}{2}\left(\frac{a P^{\prime}}{\Delta}-a \right), P^{\prime}= (r^2+ a^2)$ and  the Weierstra$\ss$ invariants take the form
\begin{eqnarray}
g_2&=&\frac{1}{12}(\alpha+\beta)^2-\frac{{\cal Q}\alpha}{4} \nonumber \\
g_3&=&\frac{1}{216}(\alpha+\beta)^3-\frac{{\cal Q}\alpha^2}{48}-\frac{{\cal Q}
\alpha\beta}{48}
\end{eqnarray}

\subsection{Exact solution for spherical  polar null geodesics with a vanishing cosmological constant}
\label{SPNULL}



In terms of the original variables, the exact solution for 
the polar orbit of the photon ($\Phi=0$) takes the form 
\begin{equation}
\wp(\phi+\epsilon)=\frac{\alpha^{\prime\prime}}{4}\cos^2 \theta-
\frac{1}{12}(\alpha^{\prime\prime}+\beta^{\prime\prime})
\label{PolikoPhos}
\end{equation}
where $\alpha^{\prime\prime}:=\alpha^{\prime}/A^{\prime 2}=
\frac{\alpha}{a^2 A^{\prime 2}}=-\frac{1}{A^{\prime 2}},\;
\beta^{\prime\prime}:=\beta^{\prime}/A^{\prime 2}=
\frac{\cal Q}{a^2 A^{\prime 2}} ,\;{\cal Q}^{\prime\prime}=
{\cal Q}^{\prime}/A^{\prime 2}=\frac{\cal Q}{a^2 A^{\prime 2}}$. Also $A^{\prime}$ is given by the 
expression
\begin{equation}
A^{\prime}:=\frac{-\frac{GMr}{c^2 a^2}}{\frac{r^2}{a^2}+1-\frac{2G Mr}{c^2 a^2}}
\end{equation}
Equation (\ref{PolikoPhos}) represents the first exact solution of a 
spherical polar photonic orbit assuming a zero cosmological constant, 
in closed analytic form, in terms of the Weierstra$\ss$ Jacobi modular 
form of weight 2.

The Weierstra$\ss$ invariants are given by
\begin{eqnarray}
g_2^{\prime\prime}&=&\frac{(\alpha^{\prime\prime}+\beta^{\prime\prime})^2}{
12}-\frac{{\cal Q}^{\prime\prime}\alpha^{\prime\prime}}{4} \nonumber \\
&=&\frac{1}{12}\frac{(-a^2+{\cal Q})^2}{a^4 A^{\prime 4}}+
\frac{\cal Q}{4 a^2 A^{\prime 4}} \nonumber \\
g_3^{\prime\prime}&=&\frac{\left(\alpha^{\prime\prime}+\beta^{\prime\prime}
\right)^3}{216}-\frac{{\cal Q}^{\prime\prime}\alpha^{\prime\prime 2}}{48}-
\frac{{\cal Q}^{\prime\prime}\alpha^{\prime\prime}\beta^{\prime\prime}}{48}
\nonumber \\
&=&\frac{1}{432 a^6 A^{\prime 6}}\left[
-2 a^6-3 a^4 {\cal Q}+3 a^2 {\cal Q}^2+2 {\cal Q}^3\right] \nonumber 
\end{eqnarray}

The sign of the discriminant $\Delta^c$ ($\Delta^c=g_2^3-27g_3^2$) 
 determines the 
roots  of the elliptic curve: $\Delta^c>0$, corresponds
to three real roots  while for $\Delta^c<0$ two roots are complex 
conjugates and the third is real. In the degenerate case $\Delta^c=0$, 
(where at least two roots coincide) the elliptic curve becomes singular and 
the solution is not given by 
modular functions. 
The analytic expressions for  the three roots of the cubic, 
which can be obtained by applying the algorithm of Tartaglia and 
Cardano \cite{TARTACARDA}, are given by

\begin{eqnarray}
e_1 &=& \frac{\left( a^2 +2 {\cal Q}\right)\Delta^2}{12 a^2  r^2 (GM/c^2)^2} \nonumber \\
e_2 &=& \frac{\left( a^2 -{\cal Q}\right)\Delta^2}{12 a^2  r^2(GM/c^2)^2} \nonumber \\
e_3 &=& -\frac{\left(2 a^2 +{\cal Q}\right)\Delta^2}{12 a^2  r^2 (GM/c^2)^2} \nonumber \\
\end{eqnarray}

Since we are assuming spherical orbits, there are two conditions from 
the vanishing of the polynomial $R(r)$ and its first derivative \footnote{
These orbits are unstable since $\frac{d^2 R}{dr^2}\Bigl|_{r=r_f}>0$. However, they represent interesting 
new possible types of motion in the Kerr spacetime. They represent a non-trivial generalisation of the unstable circular closed orbit (photonsphere) in the 
Schwarzschild black hole.}. Implementing 
these two conditions, expressions for the parameter $\Phi$ and Carter's constant ${\cal Q}$ are obtained 

\begin{eqnarray}
\Phi&=&\frac{a^2+r^2}{a},{\cal Q}=-\frac{r^4}{a^2} \nonumber \\
\Phi&=&\frac{a^2 \frac{ G M}{c^2}+a^2 r-3 \frac{G M}{c^2} r^2+r^3}{a (\frac{GM}{c^2}-r)},{\cal Q}=-\frac{r^3(-4 a^2 \frac{G M}{c^2}+r (\frac{-3 G M}{c^2}+r)^2)}{a^2 (\frac{G M}{c^2}-r)^2} \nonumber \\
\label{CON}
\end{eqnarray}
However, only the second solution is physical \cite{Teo}.

The two half-periods $\omega$ and $\omega^{\prime}$ are given by the 
following Abelian integrals (for $\Delta^c>0$) \cite{WHITAKKER}:
\begin{equation}
\omega=\int_{e_1}^{\infty}\frac{dt}{\sqrt{4t^3-g_2t-g_3}}, 
\;\;\;\omega^{\prime}=i\int_{-\infty}^{e_3}\frac{dt}{\sqrt{-4t^3+g_2t+g_3}}
\end{equation}
The values of the Weierstra$\ss$ function at the half-periods 
are 
the three roots of the cubic.
For positive discriminant $\Delta^c$ one half-period is real while the 
second is imaginary \footnote{We organize the roots as:$\;\;e_1>e_2>e_3.$}.
The period ratio is defined as $\tau=\frac{\omega^{\prime}}{\omega}$.

An alternative expression for the real half-period $\omega$ of 
the Weierstra$\ss$ function is:
$\omega=\frac{1}{\sqrt{e_1-e_3}}\frac{\pi}{2}F(\frac{1}{2},\frac{1}{2},1,
\frac{e_2-e_3}{e_1-e_3})$, where $F(\alpha,\beta,\gamma,x)$ is the hypergeometric function $1+\frac{\alpha.\beta}{1.\gamma}x+\frac{\alpha(\alpha+1)\beta(\beta+1)}{1.2.\gamma(\gamma+1)}x^2+\cdots$.

Thus \footnote{Yet another equivalent representation for $\omega$ is 
$\omega=\frac{a \pi}{\sqrt{{\cal Q}}}\frac{\frac{GMr}{c^2 a^2}}{
\frac{r^2}{a^2}+1-\frac{2GMr}{c^2a^2}}F\left(\frac{1}{2},\frac{1}{2},1,
 \frac{-a^2}{\cal Q}\right)$.}
\begin{equation}
\omega=\frac{2}{\sqrt{\frac{(a^2+{\cal Q})(a^2+(-2+r) r)^2}{a^2 r^2}}}\frac{\pi}{2}
F(\frac{1}{2},\frac{1}{2},1,\frac{a^2}{a^2+{\cal Q}})
\end{equation}

After a complete oscillation in latitude, the angle of longitude, 
which determines the amount of dragging for the spherical 
photon polar orbit in the general theory of 
relativity (GTR), increases by 
\begin{equation}
\Delta\phi^{{\rm GTR}}=4 \omega
\end{equation}

\begin{table}
\begin{center}
\begin{tabular}{|c|c|c|}\hline\hline
{\bf parameters} & {\bf half-period} & {\bf predicted dragging} \\

$a_{{\rm Galactic}}=0.52$ & $\omega=0.331117$ &  
$\Delta \phi^{{\rm GTR}}=1.32447$ \\
$\Phi=0,r=2.87313,{\cal Q}=25.8829$& &$ =75.88^{\circ} {\rm \;per\; revolution}=273192\frac{
{\rm arcs}}{{\rm revolution}}$ \\
\hline
$a_{{\rm Galactic}}=0.9939$ 
& $\omega=0.784737$ & 
$ \Delta \phi ^{{\rm GTR}}=3.13895$
 \\
$\Phi=0,r=2.42451,{\cal Q}=22.3842$ & &$=179.8^{\circ}{\rm
\;per\;revolution}=647455\frac{{\rm
arcs}}{{\rm revolution}} $ \\
\hline \hline
\end{tabular}
\end{center}
\caption{Predictions for frame dragging from galactic black hole for 
a photonic spherical polar orbit, for  
Kerr parameter $a_{{\rm Galactic}}=0.52\frac{GM_{{\rm BH}}}{c^2}$, 
$a_{{\rm Galactic}}=0.9939\frac{GM_{{\rm BH}}}{c^2}$,
respectively. The values of the radii are in units of $GM_{{\rm BH}}/c^2$, 
while those of Carter's constant ${\cal Q}$ in 
units of $(GM_{{\rm BH}}/c^2)^2$.
The period ratios, $\tau$, are $2.33615$i, $1.88276$i respectively.}
\label{EINSTEINPhoSphePOl}
\end{table}

We can also integrate the first and the third equation in (\ref{kerrgeo}).
Then we get
\begin{eqnarray}
\int c\; dt&=&-4\Biggl[\int_0^1 -\frac{1}{2}\frac{a^2 \left[\frac{r^2}{a^2}+1\right]^2}
{\frac{r^2}{a^2}+1-\frac{2 G M r}{c^2 a^2}}\frac{dz}{\sqrt{\cal Q}\sqrt{z}\sqrt{(1-z)}\sqrt{1-\left(\frac{-a^2}{{\cal Q}}\right)z}} \nonumber \\
&+&\int_0^1 \frac{a^2 (1-z) dz}{2 \sqrt{\cal Q}\sqrt{z}\sqrt{(1-z)}\sqrt{1-\left(\frac{-a^2}{{\cal Q}}\right)z}}\Biggr] \nonumber
\end{eqnarray}

Thus we obtain 
the following exact expression for $t$

\begin{eqnarray}
ct&=&-4\Biggl[-\frac{1}{2}\frac{a^2}{\sqrt{{\cal Q}}}\frac{\left[\frac{r^2}{a^2}+1\right]^2}
{\frac{r^2}{a^2}+1-\frac{2 G M r}{c^2 a^2}}\frac{\Gamma\left(\frac{1}{2}\right)
 \Gamma\left(\frac{1}{2}\right)}{\Gamma(1)}F\left(\frac{1}{2},\frac{1}{2},1,
\frac{-a^2}{{\cal Q}}\right) \nonumber \\
&+&\frac{a^2}{2 \sqrt{{\cal Q}}} \frac{\Gamma \left(\frac{1}{2}\right)\Gamma\left(\frac{3}{2}\right)}
{\Gamma(2)}F\left(\frac{1}{2},\frac{1}{2},2,\frac{-a^2}{{\cal Q}}\right)\Biggr]
\label{timepolar}
\end{eqnarray}

Similarly, if we integrate the differential equations for $t$ and $\phi$ 
we obtain

\begin{eqnarray}
\frac{c dt}{d\phi}&=&\frac{\frac{r^2+a^2}{\Delta}P^{\prime}-a^2  \sin^2 \theta}{\frac{a}{\Delta}P^{\prime}-a} \nonumber \\
&=&a+\frac{\frac{r^2 P^{\prime}}{\Delta}}{\frac{aP^{\prime}}{\Delta}-a}+
\frac{a^2  \cos^2\theta}{\frac{aP^{\prime}}{\Delta}-a} 
\end{eqnarray}
or 
\begin{equation}
ct+{\cal E}=a\phi+\frac{r^2 P^{\prime}/\Delta}{(-2 a A^{\prime})}\phi-
\frac{4a^2 / \alpha^{\prime\prime}}{(-2 a A^{\prime})}
\left( \zeta(\phi)-\frac{1}{12}(\alpha^{\prime\prime}+\beta^{\prime \prime})\phi \right)
 \end{equation}
where we used Eq.(\ref{PolikoPhos}) and the fact that, 
$\int\wp(\phi)d\phi=-\zeta(\phi)$, where $\zeta(z)$ denotes the 
Weierstra$\ss$ zeta function. Also ${\cal E}$ denotes a constant of integration.

Assuming, that the centre of the Milky Way is a black hole and that the 
 structure of spacetime  near the region $\rm {Sgr\;A^{*}}$, is described by 
the Kerr geometry as is indicated by observations Eq.(\ref{Galaxy}), 
we determined the precise frame dragging (Lense-Thirring effect) of a null orbit with a  spherical polar
geometry.
We repeated the analysis for a value of the Kerr parameter as high as 
$a_{{\rm Galactic}}=0.9939$. 
Such high values for the angular momentum of the 
black hole, have been recently reported  from x-ray flare analysis of the 
galactic centre \cite{Porquet}.
The results are displayed in table {\ref{EINSTEINPhoSphePOl}}.
Let us also mention that in \cite{Aschenbach} it has been argued that an upper bound of $a$ is given by 
$a=0.99616$.

\subsection{Null spherical polar geodesics
with the cosmological constant, Lense-Thirring effect and Appell hypergeometric 
functions}
\label{PSPLAMBDA}

We now derive an exact expression for the amount of dragging for a photonic 
spherical polar orbit in the presence of the cosmological constant, thus 
generalizing the results of the previous section.
 After a complete 
oscillation in latitude, the angle of longitude $\Delta\phi$, which determines 
the amount of dragging for the spherical polar orbit,  is given by

\begin{eqnarray}
\Delta\phi^{{\rm GTR}}&=&-4\Biggl[\alpha_1 \int_0^1 \frac{dz}{\sqrt{f(z)}}+\beta_1\int_0^1 \frac{z dz}{
\sqrt{f(z)}}\Biggr] \nonumber \\
&=& -4 \Biggl[\alpha_1\frac{1}{\sqrt{{\cal Q}}}\frac{\Gamma(\frac{1}{2})\Gamma(\frac{1}{2})}{\Gamma(1)}
\times F_1(\frac{1}{2},\frac{1}{2},1,1,-\frac{{\cal Q} a^2 \frac{\Lambda}{3}+
\Xi^3 a^2 }{{\cal Q}},\frac{-a^2 \Lambda}{3}) \nonumber \\
&+&\beta_1\frac{1}{\sqrt{{\cal Q}}}\frac{\Gamma(\frac{3}{2})\Gamma(\frac{1}{2})}{\Gamma(2)}
\times F_1(\frac{3}{2},\frac{1}{2},1,2,-\frac{{\cal Q} a^2 \frac{\Lambda}{3}+
\Xi^3 a^2 }{{\cal Q}},\frac{-a^2 \Lambda}{3}) \Biggr]\nonumber \\
\end{eqnarray}
where $f(z)=z(1-z)({\cal Q}+z ({\cal Q} a^2 \frac{\Lambda}{3}+\Xi^3 a^2 ))(1+
\frac{a^2\Lambda}{3}z)^2$ and $\alpha_1=\frac{\Xi^2 a}{2}-\frac{1}{2}
\frac{a\Xi^2 (r^2+a^2)}{\Delta_r},\;\beta_1=-\frac{1}{2}\frac{a\Xi^2(r^2+a^2)}
{\Delta_r}\frac{a^2\Lambda}{3}$.
The function
$F_1(\alpha,\beta,\beta^{\prime},
\gamma,x,y)$ is the first of the four Appell's hypergeometric functions of 
two variables $x,y$ \cite{APPELL} \footnote{The expression $(\lambda,\kappa)=
\lambda(\lambda+1)\cdots(\lambda+\kappa-1)$, and the symbol $(\lambda,0)$ 
represents $1$.},
\begin{equation}
F_1(\alpha,\beta,\beta^{\prime},
\gamma,x,y)=\sum_{m=0}^{\infty}\sum_{n=0}^{\infty}\frac{(\alpha,m+n)(\beta,m)(\beta^{\prime},n)}
{(\gamma,m+n)(1,m)(1,n)}x^m y^n
\end{equation}
which admits the following integral representation
\begin{equation}
\int_0^1 u^{\alpha-1}(1-u)^{\gamma-\alpha-1}(1-u x)^{-\beta}
(1-uy)^{-\beta^{\prime}}du=\frac{\Gamma(\alpha)\Gamma(\gamma-\alpha)}{
\Gamma(\gamma)}F_1 (\alpha,\beta,\beta^{\prime},\gamma,x,y)
\end{equation}
The double series converges when $|x|<1$ and $|y|<1$. The above Euler integral 
representation is valid for ${\rm Re}(\alpha)>0$ and ${\rm Re}(\gamma-\alpha)>0$. Also $\Gamma(p)=\int_0^{\infty}x^{p-1}e^{-x}dx$ denotes the gamma function.

For a zero cosmological constant ($\Lambda=0,\beta_1=0$) and 
we obtain the correct limit,
\begin{equation}
\Delta\phi^{{\rm GTR}}=-4\frac{\alpha_1\pi}{\sqrt{{\cal Q}}}F(\frac{1}{2},\frac{1}{2},1,\frac{-a^2 }{{\cal Q}})\end{equation}
where $\alpha_1=-\frac{aG Mr}{c^2\Delta}$, in the limit of a vanishing cosmological constant.

We can also obtain an exact expression for time. After a quarter of an oscillation in latitude the time elapses as 
\begin{eqnarray}
c\;t&=&\int_0^1\frac{(\gamma_1+\delta_1 z)dz}{\sqrt{f(z)}} \nonumber \\
&=&\frac{\gamma_1}{\sqrt{{\cal Q}}}\pi F_1\left(\frac{1}{2},\frac{1}{2},1,1,
-\frac{{\cal Q} a^2 \frac{\Lambda}{3}+\Xi^3 a^2}{{\cal Q}},\frac{-a^2\Lambda}{3}\right) \nonumber \\
&+&\frac{\delta_1}{\sqrt{{\cal Q}}}\frac{\pi}{2}F_1\left(\frac{3}{2},\frac{1}{2},
1,2,-\frac{{\cal Q} a^2 \frac{\Lambda}{3}+\Xi^3 a^2}{{\cal Q}},\frac{-a^2\Lambda}{3}\right) \nonumber \\
\end{eqnarray}

In the limit $\Lambda=0$, $\gamma_1=-\frac{1}{2}\frac{(r^2+a^2)^2}{
\Delta}+\frac{a^2}{2},\delta_1=-\frac{a^2}{2}$, and  Eq.(\ref{timepolar}) is recovered.

The conditions from the vanishing of the polynomial $R$ and its first derivative result in equations which generalize (\ref{CON}) and are 
provided in Appendix C.

\section{Non-spherical polar null geodesics}
\label{NSP}

In this case, the relevant differential equation for the calculation of 
deviation angle $\Delta\phi$  of light from the rotating black hole  
(or rotating 
central mass) is the following
\begin{equation}
\frac{d\phi}{dr}=\frac{2 a G M r}{c^2 \Delta}\frac{1}{\sqrt{R}}
\label{Framenspolar}
\end{equation}
where the quartic polynomial $R(r)$ is given by the 
expression
\begin{equation}
R=r^4+r^2 (a^2-{\cal Q})+\frac{2G M r}{c^2}(a^2+{\cal Q})-a^2 {\cal Q}
\end{equation}
Expressing the roots of $\Delta$ as $r_{+},r_{-}$ which are the locations of 
the event horizons of the black hole,
and using partial fractions we derive the expression
\begin{eqnarray}
\frac{d\phi}{dr}&=&\frac{A_{+}^P}{(r-r_{+})\sqrt{R}}+
\frac{A_{-}^P}{(r-r_{-})\sqrt{R}} \nonumber \\
&=&\frac{A_{+}^P}{(r-r_{+})\sqrt{(r-\alpha)(r-\beta)(r-\gamma)(r-\delta)}}
\nonumber \\
&+& 
\frac{A_{-}^P}{(r-r_{-})\sqrt{(r-\alpha)(r-\beta)(r-\gamma)(r-\delta)}} \nonumber
\end{eqnarray}
where $A_{\pm}^P$ are given by the equations
\begin{equation}
A_{\pm}^P=\pm\frac{2 a G M r_{\pm}}{c^2 (r_{+}-r_{-})}
\end{equation}
Also the radii of the event horizons are located at 
\begin{equation}
r_{\pm}=\frac{GM}{c^2}\pm \sqrt{\left(\frac{GM}{c^2}\right)^2-a^2}
\end{equation}
In order to calculate the angle of deflection we need to integrate the above equation from the distance of closest approach (e.g. from the maximum positive 
root of the quartic) to infinity.
 We denote the roots of the quartic by 
$\alpha,\beta,\gamma,\delta,\;\;\alpha>\beta>\gamma>\delta$. 
Thus $\Delta\phi=2\int_{\alpha}^{\infty}$.
We organize all roots in ascending order of magnitude as follows, \footnote{
We have the correspondence $\alpha_{\mu+1}=\alpha,\alpha_{\mu+2}=
\beta,\alpha_{\mu-1}=r_{+}=\alpha_{\mu-2},\alpha_{\mu-3}=\gamma,
\alpha_{\mu}=\delta$.} 
\begin{equation}
\alpha_{\mu}>\alpha_{\nu}>\alpha_i>\alpha_{\rho}
\end{equation}
where $\alpha_{\mu}=\alpha_{\mu+1},\alpha_{\nu}=\alpha_{\mu+2},
\alpha_{\rho}=\alpha_{\mu}$ and $\alpha_i=\alpha_{\mu-i},i=1,2,3$
and we have that $\alpha_{\mu-1}\geq\alpha_{\mu-2}>\alpha_{\mu-3}$.
By applying the transformation 
\begin{equation}
r=\frac{\omega z \alpha_{\mu+2}-\alpha_{\mu+1}}{\omega z-1}
\end{equation}
or equivalently 
\begin{equation}
z=\left(\frac{\alpha_{\mu}-\alpha_{\mu+2}}{\alpha_{\mu}-\alpha_{\mu+1}}
\right)\left(\frac{r-\alpha_{\mu+1}}{r-\alpha_{\mu+2}}\right)
\end{equation}
where 
\begin{equation}
\omega:=\frac{\alpha_{\mu}-\alpha_{\mu+1}}{\alpha_{\mu}-\alpha_{\mu+2}}
\end{equation}
we can bring our integrals into the familiar integral representation 
of Lauricella's $F_D$ and Appell's hypergeometric function $F_1$ of three and two variables respectively.
Indeed, we derive
\begin{eqnarray}
\Delta\phi&=&2\Biggl[\int_0^{1/\omega}\frac{-A_{+}^P\omega(\alpha_{\mu+1}-
\alpha_{\mu+2})}{H^{+}}\frac{dz}{\sqrt{z(1-z)}(1-\kappa_{+}^2 z) \sqrt{1-
\mu^2 z}} \nonumber \\
&+&\int_0^{1/\omega}\frac{A_{+}^P\omega^2(\alpha_{\mu+1}-
\alpha_{\mu+2})}{H^{+}}\frac{zdz}{\sqrt{z(1-z)}(1-\kappa_{+}^2 z) \sqrt{1-
\mu^2 z}} \nonumber \\
&+&\int_0^{1/\omega}\frac{-A_{-}^P\omega(\alpha_{\mu+1}-
\alpha_{\mu+2})}{H^{-}}\frac{dz}{\sqrt{z(1-z)}(1-\kappa_{-}^2 z) \sqrt{1-
\mu^2 z}} \nonumber \\
&+&\int_0^{1/\omega}\frac{A_{-}^P\omega^2(\alpha_{\mu+1}-
\alpha_{\mu+2})}{H^{-}}\frac{zdz}{\sqrt{z(1-z)}(1-\kappa_{-}^2 z) \sqrt{1-
\mu^2 z}}\Biggr] \nonumber \\
\end{eqnarray}
where the 
 moduli $\kappa^2_{\pm},\mu^2$ are 
\begin{eqnarray}
\kappa^2_{\pm}&=&\left(\frac{\alpha_{\mu}-\alpha_{\mu+1}}{\alpha_{\mu}-\alpha_{\mu+2}}\right)\left(
\frac{\alpha_{\mu+2}-\alpha_{\mu-1}^{\pm}}{\alpha_{\mu+1}-\alpha_{\mu-1}^{\pm}}\right)
\nonumber \\
\mu^2&=&\left(\frac{\alpha_{\mu}-\alpha_{\mu+1}}{\alpha_{\mu}-\alpha_{\mu+2}}\right)\left(
\frac{\alpha_{\mu+2}-\alpha_{\mu-3}}{\alpha_{\mu+1}-\alpha_{\mu-3}}\right)
\nonumber \\
\end{eqnarray}
Also
\begin{equation}
H^{\pm}=\sqrt{\omega}(\alpha_{\mu+1}-\alpha_{\mu+2})
(\alpha_{\mu+1}-\alpha_{\mu-1}^{\pm})\sqrt{\alpha_{\mu+1}-\alpha_{\mu}}
\sqrt{\alpha_{\mu+1}-\alpha_{\mu-3}}
\end{equation}
and $\alpha_{\mu-1}^{\pm}=r_{\pm}$.
By defining a new variable $z^{\prime}:=\omega z$ we can express the 
angle $\Delta\phi$ in terms of Lauricella's hypergeometric function $F_D$,
\begin{eqnarray}
\Delta\phi^{{\rm GTR}}&=&2\Biggl[\frac{-2 A_{+}^P\sqrt{\omega}(\alpha_{\mu+1}-
\alpha_{\mu+2})}{H^{+}}F_D\left(\frac{1}{2},\frac{1}{2},1,\frac{1}{2},
\frac{3}{2},\frac{1}{\omega},\kappa^{\prime 2}_+,\mu^{\prime 2}\right) \nonumber \\
&+&\frac{A_{+}^P\sqrt{\omega}(\alpha_{\mu+1}-
\alpha_{\mu+2})}{H^{+}}F_D\left(\frac{3}{2},\frac{1}{2},1,\frac{1}{2},
\frac{5}{2},\frac{1}{\omega},\kappa^{\prime 2}_+,\mu^{\prime 2}\right)
\frac{\Gamma(3/2)\Gamma(1)}{\Gamma(5/2)} \nonumber \\       
&+&\frac{-2A_{-}^P\sqrt{\omega}(\alpha_{\mu+1}-
\alpha_{\mu+2})}{H^{-}}F_D\left(\frac{1}{2},\frac{1}{2},1,\frac{1}{2},
\frac{3}{2},\frac{1}{\omega},\kappa^{2\prime}_-,\mu^{\prime 2}\right) \nonumber \\
&+&\frac{A_{-}^P\sqrt{\omega}(\alpha_{\mu+1}-
\alpha_{\mu+2})}{H^{-}}F_D\left(\frac{3}{2},\frac{1}{2},1,\frac{1}{2},
\frac{5}{2},\frac{1}{\omega},\kappa^{\prime 2}_-,\mu^{\prime 2}\right)
\frac{\Gamma(3/2)\Gamma(1)}{\Gamma(5/2)}\Biggr] \nonumber \\  
\label{PolarBending}
\end{eqnarray}
where the variables of the function $F_D$ are given in terms 
of the roots of the quartic and the radii of the event horizons 
by the expressions
\begin{eqnarray}
\frac{1}{\omega}&=&\frac{\alpha_{\mu}-\alpha_{\mu+2}}{\alpha_{\mu}-\alpha_{\mu+1}}=\frac{\delta-\beta}{\delta-\alpha} \nonumber \\
\kappa^{\prime 2}_{\pm}&=&\frac{\alpha_{\mu+2}-\alpha_{\mu-1}^{\pm}}{
\alpha_{\mu+1}-\alpha_{\mu-1}^{\pm}}=\frac{\beta-r_{\pm}}{\alpha-r_{\pm}}
\nonumber \\
\mu^{\prime 2}&=&\frac{\alpha_{\mu+2}-\alpha_{\mu-3}}{\alpha_{\mu+1}-
\alpha_{\mu-3}}=\frac{\beta-\gamma}{\alpha-\gamma} \nonumber \\
\end{eqnarray}
An equivalent expression is as follows
\begin{eqnarray}
\Delta\phi^{{\rm GTR}}&=&2\Biggl[
\frac{-2 A_{+}^P\sqrt{\omega}(\alpha_{\mu+1}-
\alpha_{\mu+2})}{H^{+}}F_D\left(\frac{1}{2},\frac{1}{2},1,\frac{1}{2},
\frac{3}{2},\frac{1}{\omega},\kappa^{\prime 2}_+,\mu^{\prime 2}\right) 
\nonumber \\
&+&\frac{A_{+}^P\sqrt{\omega}(\alpha_{\mu+1}-
\alpha_{\mu+2})}{H^{+}}\Biggl(-\frac{1}{\kappa^{\prime 2}_{+}}
F_1\left(\frac{1}{2},\frac{1}{2},\frac{1}{2},\frac{3}{2},\frac{1}{\omega},
\mu^{\prime 2}\right) 2 \nonumber \\
&+&\frac{1}{\kappa^{\prime 2}_{+}}F_D\left(\frac{1}{2},
\frac{1}{2},1,\frac{1}{2},\frac{3}{2},\frac{1}{\omega},\kappa^{\prime 2}_{+},
\mu^{\prime 2}\right) 2 \Biggr)\nonumber \\
&+&\frac{-2A_{-}^P\sqrt{\omega}(\alpha_{\mu+1}-
\alpha_{\mu+2})}{H^{-}}F_D\left(\frac{1}{2},\frac{1}{2},1,\frac{1}{2},
\frac{3}{2},\frac{1}{\omega},\kappa^{\prime 2}_-,\mu^{\prime 2}\right) \nonumber \\
&+&\frac{A_{-}^P\sqrt{\omega}(\alpha_{\mu+1}-
\alpha_{\mu+2})}{H^{-}}
\Biggl(-\frac{1}{\kappa^{\prime 2}_{-}}
F_1\left(\frac{1}{2},\frac{1}{2},\frac{1}{2},\frac{3}{2},\frac{1}{\omega},
\mu^{\prime 2}\right) 2 \nonumber \\
&+&\frac{1}{\kappa^{\prime 2}_{-}}F_D\left(\frac{1}{2},
\frac{1}{2},1,\frac{1}{2},\frac{3}{2},\frac{1}{\omega},\kappa^{\prime 2}_{-},
\mu^{\prime 2}\right) 2 \Biggr)\Biggr]\nonumber \\
\label{PolarDeflection}
\end{eqnarray}
In going from Eq.(\ref{PolarBending}) to Eq.(\ref{PolarDeflection}) we 
used the identity which is proven in Appendix \ref{Appell1}
\begin{eqnarray}
& &F_D\left(\frac{3}{2},\frac{1}{2},1,\frac{1}{2},\frac{5}{2},
\frac{1}{\omega},\kappa^{\prime 2}_+,\mu^{\prime 2}\right)\frac{\Gamma(3/2)}{
\Gamma(5/2)} \nonumber \\
&=&-\frac{1}{\kappa^{\prime 2}_{+}}F_1\left(\frac{1}{2},\frac{1}{2},\frac{1}{2},
\frac{3}{2},\frac{1}{\omega},\mu^{\prime 2}\right)\frac{\Gamma(1/2)}{\Gamma(3/2)}\nonumber \\
&+&\frac{1}{\kappa^{\prime 2}_{+}}F_D\left(\frac{1}{2},\frac{1}{2},1,
\frac{1}{2},\frac{3}{2},\frac{1}{\omega},\kappa^{\prime 2}_+,\mu^{\prime 2}\right)\frac{\Gamma(1/2)}{\Gamma(3/2)} \nonumber \\
\label{Identity}
\end{eqnarray}

The phenomenological applications of Eq.(\ref{PolarBending}) for gravitational 
bending and lensing studies from a galactic black hole, as well as its 
generalization in the presence of the cosmological constant will be the subject of detailed investigation in a future publication.

\subsection{Exact solution of timelike spherical polar orbits with a cosmological constant}

Using the second and the fifth line of Eq.(\ref{LambdaGeo}), for $L=0$ and 
assuming a constant value for $r$, 
we obtain 
\begin{eqnarray}
\frac{d\phi}{d\theta}&=&\frac{-\frac{\Xi^2 a E}{\Delta_{\theta}}+\frac{a \Xi^2 (
r^2+a^2)E}{\Delta_r}}{\sqrt{\Theta^{\prime}}} \nonumber \\
&=&\frac{-\Xi^2 E a+B \Delta_{\theta}}{\Delta_{\theta} \sqrt{\Theta^{\prime}}}
\end{eqnarray}
where $B:=\frac{a \Xi^2 (r^2+a^2)E}{\Delta_r}$. 

Similarly, using the third and fifth line we obtain
\begin{eqnarray}
\frac{cdt}{d\theta}&=&\frac{ \Xi^2 (r^2+a^2)^2 E}{\Delta_r\sqrt{\Theta^{\prime}}}-\frac{a^2 \Xi^2 E \sin^2\theta}{\Delta_{\theta} \sqrt{\Theta^{\prime}}} \nonumber \\
&=&\frac{\Gamma \Delta_{\theta}-a^2 \Xi^2 E \sin^2\theta}{\Delta_{\theta} \sqrt{\Theta^{\prime}}}
\end{eqnarray}
and $\Gamma:=\frac{\Xi^2 (r^2+a^2)^2 E}{\Delta_r}$.

Now using the variable $z=\cos^2 \theta$, we obtain the following system 
of integral equations:
\begin{eqnarray}
\phi&=&\int \frac{\Xi^2 a E/2}{\sqrt{f(z)}}dz+\int \frac{B(1+
\frac{a^2 \Lambda}{3}z)/(-2)}{\sqrt{f(z)}}dz \nonumber \\
ct&=&\int \frac{\frac{-\Gamma}{2} (1+\frac{a^2 \Lambda}{3}z)}
{\sqrt{f(z)}}dz-\int \frac{\frac{a^2}{-2} \Xi^2 E (1-z)dz}{\sqrt{f(z)}}
\label{ABTJAI}
\end{eqnarray}
or
\begin{eqnarray}
\phi&=&\int^z \frac{(\alpha_1+\beta_1z)dz}{\sqrt{f(z)}} \nonumber \\
c\;t&=&\int^z \frac{(\gamma_1+\delta_1 z)dz}{\sqrt{f(z)}} 
\end{eqnarray}
where $f(z)=z (1-z) (Q+z (Qa^2\frac{\Lambda}{3}+\Xi^3a^2 E^2-\mu^2a^2)+
z^2 (-\mu^2a^4\frac{\Lambda}{3}))(1+a^2 \frac{\Lambda}{3} z)^2$.
Also we have defined
\begin{eqnarray}
  \alpha_1&=&\Xi^2\frac{a E}{2}-\frac{1}{2}\frac{a\Xi^2(r^2+a^2)E}{\Delta_r}
\nonumber \\
\beta_1&=&-\frac{1}{2}\frac{a\Xi^2 (r^2+a^2)E}{\Delta_r}\frac{a^2\Lambda}{3}
\nonumber \\
\gamma_1&=&-\frac{1}{2}\frac{\Xi^2 (r^2+a^2)^2E}{\Delta_r}+\frac{a^2\Xi^2E}{2}
\nonumber \\
\delta_1&=&-\frac{a^2\Lambda}{6}\frac{\Xi^2(r^2+a^2)^2 E}{\Delta_r}-
\frac{a^2\Xi^2E}{2} 
\end{eqnarray}

Equation (\ref{ABTJAI}) is a system of equations of Abelian  
integrals, whose $inversion$ in principle, involves genus-2 
Abelian-Siegelsche modular 
functions.
Indeed, this system is a particular case of Jacobi's inversion problem of hyperelliptic 
Abelian integrals of genus 2 \cite{Abel}-\cite{BAKER} (see Appendix A 
for details). Then, one can express 
$z$ as a single valued 
genus two Abelian theta function with argumenents $t,\phi$. However, 
since the polynonial $f(z)$ of sixth degree posses a double root it may 
well be that the Abelian genus-2 theta function degenerates and the 
final result can be expressed in terms of genus-1 modular functions. 

\section{Frame dragging in spherical polar timelike geodesics with a cosmological constant} 
\label{TPOLARFRAME}

After a quarter of oscillation in latitude the change of longitude is 
\begin{eqnarray}
\Delta\phi&=&\int_0^1 \frac{(\alpha_1+\beta_1 z)dz}{\sqrt{f(z)}} \nonumber \\
&=&
\frac{\alpha_1}{\sqrt{Q}}\frac{\Gamma(\frac{1}{2})(\frac{1}{2})}{\Gamma(1)}
F_D(\frac{1}{2},\frac{1}{2},\frac{1}{2},1,1;
z_1^{\prime},z_2^{\prime},\frac{-a^2 \Lambda}{3})+
\frac{\beta_1}{\sqrt{Q}}\frac{\Gamma(\frac{3}{2})(\frac{1}{2})}{\Gamma(2)}
F_D(\frac{3}{2},\frac{1}{2},\frac{1}{2},1,2;
z_1^{\prime},z_2^{\prime},\frac{-a^2 \Lambda}{3}) \nonumber
\end{eqnarray}
where $z_i^{\prime}=1/z_i$ and 
\begin{eqnarray}
z_1 &=&  -\frac{-3 a^2 E^2 \Xi ^3+3
   a^2-a^2 Q \Lambda -\sqrt{12 Q \Lambda  a^4+\left(3 a^2
   E^2 \Xi ^3-3 a^2+a^2 Q \Lambda \right)^2}}{2
   a^4 \Lambda },\nonumber \\
z_2&=&\frac{-3 a^2
   E^2 \Xi ^3+3 a^2-a^2 Q \Lambda +\sqrt{12 Q
   \Lambda  a^4+\left(3 a^2 E^2 \Xi ^3-3 a^2+a^2
   Q \Lambda \right)^2}}{2 a^4 \Lambda } \nonumber 
\end{eqnarray}

\section{Spherical non-polar null geodesics}
\label{SNP}

In this section, we investigate spherical null geodesics with a nonzero 
value of the parameter $\Phi$.
Now one has to calculate the integral

\begin{equation}
\Delta\phi=-\frac{1}{2}\int \frac{dz \left( \frac{a}{\Delta} P-a \right)}{
\sqrt{\alpha z^3-z^2 (\alpha+\beta)+{\cal Q}z}}-\frac{1}{2}\int \frac{dz\; \Phi}{
(1-z) \sqrt{\alpha z^3-z^2 (\alpha+\beta)+{\cal Q}z}}
\label{SPHENP}
\end{equation}
where 
\begin{equation}
\alpha:=-a^2,\beta:={\cal Q}+\Phi^2,\;\;\;P=r^2+a^2-a \Phi
\end{equation}

The first integral can be brought into the Weierstra$\ss$ form with the 
invariants $g_2,g_3$ 
\begin{eqnarray}
g_2^{\prime\prime}&=&\frac{1}{12}(\alpha^{\prime\prime}+\beta^{\prime\prime})^2
-\frac{{\cal Q}^{\prime\prime}\alpha^{\prime\prime}}{4} \nonumber \\
&=&\frac{1}{12}\frac{1}{a^4 A^{\prime 4}}\left(-a^2+{\cal Q}+\Phi^2\right)^2+
\frac{{\cal Q}}{4 a^2 A^{\prime 4}} \nonumber \\
g_3^{\prime\prime}&=&\frac{1}{432 a^6 A^{\prime 6}}\left[2 (-a^2+{\cal Q}+\Phi^2)^3-9 {\cal Q} a^4+9{\cal Q}a^2 ({\cal Q} +\Phi^2)\right] \nonumber 
\end{eqnarray}
and
\begin{equation}
A^{\prime}=\frac{\frac{\Phi}{2 a}-\frac{G M r}{c^2 a^2}}{
\frac{r^2}{a^2}+1-\frac{2 G M r}{c^2 a^2}}
\end{equation}

Now let us discuss the second integral in Eq.(\ref{SPHENP}). We define a new 
variable 
\begin{equation}
u_1^{\prime} z\equiv v
\end{equation}
then 
\begin{eqnarray}
&-&\frac{1}{2}\int_0^1 \frac{dz\; \Phi}{
(1-z) \sqrt{\alpha z^3-z^2 (\alpha+\beta)+{\cal Q}z}} \nonumber \\
&=&-\int_0^{u_1^{\prime}}\frac{\Phi}{2\sqrt{{\cal Q}}\sqrt{u_1^{\prime}}}\frac{
dv}{\sqrt{v}\sqrt{1-v}(1-\chi_1 v)\sqrt{1-\chi_2 v}} \nonumber \\
&=&-\frac{\Phi}{2\sqrt{{\cal Q}}\sqrt{u_1^{\prime}}}\frac{\Gamma\left(\frac{1}{2}
\right)\Gamma\left(\frac{1}{2}
\right)}{\Gamma(1)} F_1 \left(\frac{1}{2},1,\frac{1}{2},1,\chi_1,\chi_2\right) 
\nonumber \\
&=& -\frac{\Phi}{2\sqrt{{\cal Q}}\sqrt{u_1^{\prime}}}\pi F_1\left(\frac{1}{2},1,\frac{1}{2},1,\chi_1,\chi_2\right) 
\nonumber
\end{eqnarray}
where $u_i^{\prime}=\frac{1}{u_i},i=1,2$ and 
\begin{eqnarray}
u_1&=&-\frac{-a^2+{\cal Q}+\Phi^2-\sqrt{4 a^2 {\cal Q}+(a^2-{\cal Q}-\Phi^2)^2}}{2 a^2} \nonumber \\
u_2&=&-\frac{-a^2+{\cal Q}+\Phi^2+\sqrt{4 a^2 {\cal Q}+(a^2-{\cal Q}-\Phi^2)^2}}{2 a^2} \nonumber
\end{eqnarray}
Also
\begin{equation}
\chi_1:=\frac{1}{u_1^{\prime}},\;\;\;\chi_2:=\frac{u_2^{\prime}}{u_1^{\prime}}
\end{equation}
Thus, we expressed the above integral in 
terms of Appell's first hypergeometric 
function of two variables, $ F_1(\alpha,\beta,\beta^{\prime},
\gamma,x,y)$.

After a complete oscillation in latitude, the angle of longitude, 
which determines the amount of dragging for the spherical 
non-polar photonic orbit in the General Theory of 
Relativity (GTR), is given by 
\begin{eqnarray}
\Delta\phi^{{\rm GTR}}&=&4 \frac{\pi}{2 \sqrt{{\cal Q}}}\frac{\left[\frac{a}{\Delta}P-a
\right]}{\sqrt{u_1^{\prime}}}F\left(\frac{1}{2},\frac{1}{2},1,\frac{
u_2^{\prime}}{u_1^{\prime}}\right) 
+4 \frac{\Phi}{2\sqrt{{\cal Q}}\sqrt{u_1^{\prime}}}\pi F_1\left(\frac{1}{2},1,\frac{1}{2},1,\chi_1,\chi_2\right) \nonumber \\
&=&4 \frac{\pi}{2 \sqrt{{\cal Q}}}\frac{1}{\sqrt{u_1^{\prime}}}
\frac{a \left(\frac{-\Phi}{a}+\frac{2GM r}{c^2a^2}\right)}
{\frac{r^2}{a^2}+1- \frac{2 G M r}{c^2 a^2}}F\left(\frac{1}{2},
\frac{1}{2},1,\frac{
u_2^{\prime}}{u_1^{\prime}}\right) 
+4 \frac{\Phi}{2\sqrt{{\cal Q}}\sqrt{u_1^{\prime}}}\pi F_1\left(\frac{1}{2},1,\frac{1}{2},1,\chi_1,\chi_2\right) \nonumber \\
\label{LTNPOLAR}
\end{eqnarray}

\begin{table}
\begin{center}
\begin{tabular}{|c|c|}\hline\hline
{\bf parameters} & {\bf predicted dragging} \\

$\Phi=1,{\cal Q}=22.693, r=2.7452$ & $\Delta \phi^{{\rm GTR}}=7.72736=442.7^{\circ} {\rm \;per\; revolution}=1.59\times 10^6\frac{
{\rm arcs}}{{\rm revolution}}$ \\
$\Phi=-1,{\cal Q}=26.9984, r=2.99523$ 
& $ \Delta \phi ^{{\rm GTR}}=-5.0551=-289.6^{\circ}{\rm
\;per\;revolution}=-1.04\times 10^6\frac{{\rm
arcs}}{{\rm revolution}}$ \\
$\Phi=-3,{\cal Q}=23.0508, r=3.2239$ 
& $ \Delta \phi ^{{\rm GTR}}=-5.2014=-298^{\circ}{\rm
\;per\;revolution}=-1.07\times 10^6\frac{{\rm
arcs}}{{\rm revolution}}$ \\
\hline \hline
\end{tabular}
\end{center}
\caption{Predictions for frame dragging from a galactic black hole, with 
Kerr parameter $a_{{\rm Galactic}}=0.52\frac{GM_{{\rm BH}}}{c^2}$, for  different 
values of photon angular momentum and Carter's Constant. The values of the radii and $\Phi$ are in units of $GM_{{\rm BH}}/c^2$, 
while Carter's constant ${\cal Q}$ in units 
of $(GM_{{\rm BH}}/c^2)^2$.}
\label{EINSTEIN1photon}
\end{table}

\begin{table}
\begin{center}
\begin{tabular}{|c|c|}\hline\hline
{\bf parameters} & {\bf predicted dragging} \\

$\Phi=1,{\cal Q}=16.1443, r=2.02083$ & $\Delta \phi^{{\rm GTR}}=10.7355=615^{\circ} {\rm \;per\; revolution}=2.2\times 10^6\frac{
{\rm arcs}}{{\rm revolution}}$ \\
$\Phi=-1,{\cal Q}=25.8865, r=2.73783$ 
& $ \Delta \phi ^{{\rm GTR}}=-3.73503=-214^{\circ}{\rm
\;per\;revolution}=-770405\frac{{\rm
arcs}}{{\rm revolution}}$ \\
$\Phi=-3,{\cal Q}=25.8628, r=3.23713$ 
& $ \Delta \phi ^{{\rm GTR}}=-4.32779=-247.9^{\circ}{\rm
\;per\;revolution}=-892671\frac{{\rm
arcs}}{{\rm revolution}}$ \\
\hline \hline
\end{tabular}
\end{center}
\caption{Predictions for frame dragging from a galactic black hole, with 
Kerr parameter $a_{{\rm Galactic}}=0.9939\frac{GM_{{\rm BH}}}{c^2}$, for  different 
values of photon angular momentum and Carter's Constant. The values of the radii and $\Phi$ are in units of $GM_{{\rm BH}}/c^2$, 
while Carter's constant ${\cal Q}$ in units 
of $(GM_{{\rm BH}}/c^2)^2$.}
\label{EINSTEIN2photon}
\end{table}
Orbits with $\Delta\phi^{\rm GTR}>0$ are called $prograde$ and those
with $\Delta\phi^{\rm GTR}<0$ are called $retrogade$.

As before we can obtain an exact expression for time. After a quarter of 
oscillation in latitude 
\begin{eqnarray}
\int c\;dt &=& -[-\int_0^1 \frac{dz}{2\sqrt{z}}\frac{
\frac{(r^2+a^2)}{\Delta}\left[(r^2+a^2)-a\Phi\right]}{
\sqrt{\alpha z^2-z(\alpha+\beta)+{\cal Q}}} \nonumber \\
&+&\frac{dz}{2\sqrt{z}}\frac{a[a(1-z)-\Phi]}{\sqrt{\alpha z^2-z(\alpha+\beta)+{\cal Q}}}] \nonumber \\
&=&-[\left[-\frac{1}{2}\frac{r^2+a^2}{\Delta}[(r^2+a^2)-a \Phi]-a\frac{\Phi}{2}\right]\frac{1}{\sqrt{{\cal Q}}}\frac{1}{\sqrt{u_1^{\prime}}}F\left(\frac{1}{2},\frac{1}{2},1,\frac{u_2^{\prime}}{u_1^{\prime}}\right)\pi \nonumber \\
&+&\frac{a^2}{2\sqrt{{\cal Q}}}\frac{\Gamma\left(\frac{1}{2}\right)\Gamma(2)}{
\Gamma\left(\frac{5}{2}\right)}F_1\left(\frac{1}{2},\frac{1}{2},
\frac{1}{2},\frac{5}{2},u_1^{\prime},u_2^{\prime}\right) ]
\label{timenpolar}
\end{eqnarray}
The above equation has the correct limit for $\Phi=0$ and reproduces 
the corresponding exact expressiion, Eqn.(\ref{timepolar}) for 
spherical null polar geodesics.  Indeed, for $\Phi=0$, 
$\frac{u_2^{\prime}}{u_1^{\prime}}=\frac{-a^2}{{\cal Q}},u_1^{\prime}=1$, and the 
Appell function has the following limit:
\begin{equation}
\frac{\Gamma\left(\frac{1}{2}\right)\Gamma(2)}{\Gamma\left(\frac{1}{2}\right)}
F_1 \left(\frac{1}{2},\frac{1}{2},
\frac{1}{2},\frac{5}{2},1,\frac{-a^2}{{\cal Q}}\right) =
\frac{\Gamma\left(\frac{1}{2}\right)\Gamma\left(\frac{3}{2}\right)}
{\Gamma(2)}F\left(\frac{1}{2},\frac{1}{2},2,\frac{-a^2}{{\cal Q}}\right)
\end{equation}
Assuming that the centre of the Milky Way is a rotating black hole and that 
the structure of spacetime near the region Sgr ${\rm A^{*}}$ is described by the 
Kerr geometry, we determined the precise frame dragging of a null orbit 
with a spherical non-polar geometry. The results are displayed in tables 
2 and 3.

\section{Frame dragging in spherical non-polar null geodesics with a 
cosmological constant}
\label{SNPLAMBDA}

Including the contribution from the cosmological constant,
the relevant differential equation is
\begin{equation}
\frac{d\phi}{d\theta}=\frac{\frac{-a\Xi^2}{\Delta_{\theta}}
+\frac{\Phi\Xi^2}{\Delta_{\theta}\sin^2 \theta}}{\sqrt{\Theta^{\prime}}}+
\frac{B^{\prime}}{\Delta_r\sqrt{\Theta^{\prime}}}
\end{equation}
where  $B^{\prime}:=\Xi^2 a P$.

Introducing the variable $z$, we find that
the angle of longitude that measures the frame-dragging of a spherical 
non-polar photon orbit,
after a complete oscillation in latitude is given by 
the exact expression

\begin{eqnarray}
\Delta\Phi^{\rm GTR}&=&-4\Biggl[-\frac{1}{2}\frac{\Phi\Xi^2}{
\sqrt{{\cal Q}}}
\frac{1}{\sqrt{u_1^{\prime}}}\frac{\pi}{1}F_D\left(\frac{1}{2},
1,1,\frac{1}{2},1,\frac{-a^2 \Lambda}{3u_1^{\prime}},\frac{1}{u_1^{\prime}},
\frac{u_2^{\prime}}{u_1^{\prime}}\right) \nonumber \\
&+&\frac{1}{2}\frac{a\Xi^2}{\sqrt{{\cal Q} u_1^{\prime}}}
\pi F_1\left(\frac{1}{2},1,\frac{1}{2},1,-\frac{a^2\Lambda}{3
u_1^{\prime}},\frac{u_2^{\prime}}{u_1^{\prime}}\right) \nonumber \\
&-&\frac{B^{\prime}}{\Delta_r}\frac{1}{\sqrt{{\cal Q} u_1^{\prime}}}\frac{\pi}{2}F\left(\frac{1}{2},\frac{1}{2},1,\frac{u_2^{\prime}}{
u_1^{\prime}}\right)\Biggr] \nonumber \\
\label{NonPOLARLAMBDA}
\end{eqnarray}
where $u_i^{\prime}=1/u_i,i=1,2$ and the roots $u_i$ are 
\begin{eqnarray}
u_1&=&\frac{\Lambda  \Xi ^2 a^4-2 \Lambda  \Xi
   ^2 \Phi  a^3+3 \Xi ^2 a^2+\Lambda  \Xi ^2 \Phi ^2 a^2+{\cal Q} \Lambda  a^2-3 \Xi ^2
   \Phi ^2-3 {\cal Q}+\sqrt{H_{\Lambda}}}{2 a^2 \left(\left(\Lambda  a^2-2 \Lambda  \Phi  a+\Lambda
    \Phi ^2+3\right) \Xi ^2+{\cal Q} \Lambda \right)} \nonumber \\
u_2&=&\frac{\Lambda  \Xi ^2 a^4-2 \Lambda  \Xi ^2 \Phi  a^3+3 \Xi ^2
   a^2+\Lambda  \Xi ^2 \Phi ^2 a^2+{\cal Q} \Lambda  a^2-3 \Xi ^2 \Phi ^2-3 {\cal Q}-\sqrt{H_{\Lambda}}}{2 a^2
   \left(\left(\Lambda  a^2-2 \Lambda  \Phi  a+
\Lambda  \Phi ^2+3\right) \Xi ^2+{\cal Q}
   \Lambda \right)} \nonumber \\
\end{eqnarray}
and 
\begin{eqnarray}
H_{\Lambda}&:=&12 {\cal Q}
   \left(\left(\Lambda  a^2-2 
\Lambda  \Phi  a+\Lambda  \Phi ^2+3\right) \Xi ^2+{\cal Q}
   \Lambda \right) a^2 \nonumber \\
&+&\left((a-\Phi ) \left(\Lambda  a^3-\Lambda  \Phi  a^2+3
   a+3 \Phi \right) \Xi ^2+{\cal Q} \left(a^2 \Lambda -3\right)\right)^2 \nonumber \\
\end{eqnarray}

 Equation (\ref{NonPOLARLAMBDA}) involves 
three hypergeometric functions, Gau$\ss$'s $F$, Appell's $F_1$ and 
Lauricella's $F_D$. For vanishing 
cosmological constant the above expression reduces exactly to 
(\ref{LTNPOLAR}).

Similarly the time elapses after a complete oscillation in latitude is 

\begin{eqnarray}
ct&=&-4\Biggl[-\frac{1}{2}\frac{\Xi^2(r^2+a^2)[r^2+a^2-a\Phi]}{\Delta_r}
\frac{1}{\sqrt{{\cal Q}u_1^{\prime}}}F\left(\frac{1}{2},\frac{1}{2},1,
\frac{u_2^{\prime}}{u_1^{\prime}}\right)\pi \nonumber \\
&-&\frac{1}{2}\frac{a\Phi\Xi^2}{\sqrt{{\cal Q}u_1^{\prime}}}F_1\left(
\frac{1}{2},1,\frac{1}{2},1,\frac{a^2\Lambda}{3u_1^{\prime}},
\frac{u_2^{\prime}}{u_1^{\prime}}\right)\pi \nonumber \\
&+&\frac{a^2 \Xi^2}{2\sqrt{{\cal Q}u_1^{\prime}}}F_D\left(\frac{1}{2},-1,
1,\frac{1}{2},1,\frac{1}{u_1^{\prime}},\frac{a^2\Lambda}{3u_1^{\prime}},
\frac{u_2^{\prime}}{u_1^{\prime}}\right)\pi\Biggr] \nonumber \\
\end{eqnarray}

\section{Spherical timelike geodesics with $L \not = 0 $}
\label{LNZSPHE}

The relevant equation for integration is \cite{KraniotisKerr}
\begin{equation}
d\phi=-\frac{1}{2}\frac{dz}{\sqrt{z^3 \alpha-z^2 (\alpha+\beta)+Q z}}
\times \left\{\frac{aP}{\Delta}-aE+\frac{L}{1-z}\right\}
\end{equation}
where 
\begin{equation}
\alpha=a^2 (1-E^2),\;\;\;\beta=Q+L^2
\end{equation}
and $P$ is provided from Eq.(\ref{GenikiP}) \footnote{The extreme black hole 
$a=1$ solutions for $L\not=0$ have been investigated in \cite{Wilkins}}.

Let us define:
\begin{equation}
\Pi:=\int\frac{d\xi}{\sqrt{4 \xi^3- g_2 \xi-g_3}}
\end{equation}
thus $\xi=\wp(\Pi+\epsilon)$, and $A^{\prime \prime} \int\frac{dz}{\sqrt{z^3 \alpha-
z^2 (\alpha+\beta)+Qz}}= A^{\prime \prime}\int\frac{d\xi}{\sqrt{4 \xi^3- g_2 \xi-g_3}}=
A^{\prime \prime}\Pi,\; A^{\prime \prime}:= -\frac{1}{2}\left(\frac{a P}{\Delta}-a E\right)$ and $\epsilon$ is a constant of integration.
Now
\begin{equation}
-\frac{1}{2} L \int \frac{dz}{(1-z)\sqrt{z^3 \alpha-
z^2 (\alpha+\beta)+Q z}}
\end{equation}
under the substitution (\ref{subwei}) becomes 
\begin{eqnarray}
&-&\frac{L \alpha}{8}\int \frac{d\xi}{\left(\frac{\alpha}{4}(1-
\frac{\alpha+\beta}{3\alpha})-\xi\right)\sqrt{4 \xi^3-g_2 \xi-g_3}} \nonumber \\
&=&-\frac{L \alpha}{8}\int \frac{d\xi}{\left(w-\xi\right)\sqrt{4 \xi^3-g_2 \xi-g_3}} \nonumber \\
&=&- \frac{L \alpha}{8}\int \frac{\wp^{\prime}(\Pi)d \Pi}{\left(w-
\wp(\Pi+\epsilon)\right)\sqrt{4 \wp^3(\Pi)-g_2 \;\wp(\Pi)-g_3}} \nonumber \\
&=&- \frac{L \alpha}{8}\int \frac{d\Pi}{\left(w-\wp(\Pi+\epsilon)
\right)} \nonumber \\
&=&- \frac{L \alpha}{8}\left[{\rm Log} \frac{\sigma(\Pi+\epsilon-v_0)}{
\sigma(\Pi+\epsilon+v_0)}+2 \Pi\; \zeta(v_0)\right]\times \frac{1}{\wp^{\prime}
(v_0)} \nonumber \\
\end{eqnarray}
where $w:=\frac{\alpha}{4}\left(1-\frac{\alpha+\beta}{3\alpha}\right) = \wp(v_0)$.
Also $\sigma(z)$ denotes the Weierstra$\ss$ sigma function.
Thus the equation for the orbit is given by
\begin{equation}
\phi=\int d\phi=A^{\prime \prime}\Pi- \frac{L \alpha}{8}\left[{\rm Log} \frac{\sigma(\Pi+\epsilon-v_0)}{
\sigma(\Pi+\epsilon+v_0)}+2 \Pi \zeta(v_0)\right]\times \frac{1}{\wp^{\prime}
(v_0)}
\end{equation}
and $\wp^{2 \prime}(v_0)=4 \wp^3(v_0)-g_2 \wp(v_0)-g_3=4 w^3-g_2 w-g_3$.
In terms of the integration constants, $w$ is given by the expression:
\begin{equation}
w=\frac{a^2(1-E^2)}{4}-\frac{a^2(1-E^2)+Q+L^2}{12}
\end{equation}

Similarly using the first and third line of Eq.(\ref{kerrgeo}), we obtain for $t$ the expression
\begin{equation}
c\; t=\frac{r^2+a^2}{\Delta}P\frac{\Pi}{-2}+\frac{a\Pi}{2}(aE-L)+
\frac{a^2E}{2}\Pi\left(-\frac{1}{3}\frac{\alpha+\beta}{\alpha}\right)+
\frac{a^2 E}{2}\frac{4}{\alpha}\zeta(\Pi)
\end{equation}

An alternative exact expression for time $t$ in terms of ordinary 
hypergeometric of one-variable $F$ and Appell's first hypergeometric
function of two variables is as follows:

\begin{eqnarray}
c\;t&=& -\frac{\pi}{2\sqrt{Q}}\frac{\left[\frac{r^2+a^2}{\Delta}P+a L\right]}{
\sqrt{u_1^{\prime}}}F\left(\frac{1}{2},\frac{1}{2},1,\frac{u_2^{\prime}}
{u_1^{\prime}}\right) \nonumber \\
&+&\frac{a^2 E}{2 \sqrt{Q}}\frac{\Gamma\left(\frac{1}{2}\right) 
\Gamma\left(\frac{1}{2}\right)}{\Gamma(1)}F_1\left(\frac{1}{2},-1,\frac{1}{2}
,1,\frac{1}{u_1^{\prime}},\frac{u_2^{\prime}}{u_1^{\prime}}\right)
\end{eqnarray}

The variables $u_i$ are given in terms of the constants of integration 
and the Kerr parameter $a$ by the expressions
\begin{eqnarray}
u_1&=&\frac{a^2 (1-E^2)+L^2+Q-\sqrt{(-a^2 (1-E^2)-L^2-Q)^2-4a^2 (1-E^2)Q}}
{2 a^2 (1-E^2)} \nonumber \\
u_2&=&\frac{a^2 (1-E^2)+L^2+Q+\sqrt{(-a^2 (1-E^2)-L^2-Q)^2-4a^2 (1-E^2)Q}}
{2 a^2 (1-E^2)} \nonumber \\
\end{eqnarray}
and $u_i^{\prime}=1/u_i,i=1,2$.

Similarly an alternative expression for the amount of dragging for timelike 
non-polar spherical orbits  
is as follows,
\begin{eqnarray}
\Delta \phi^{\rm GTR}&=&4 \frac{1}{2 \sqrt{Q}}\frac{a \left(\frac{-L}{a}+
\frac{2 G M E r}{c^2 a^2}\right)}{\frac{r^2}{a^2}+1-\frac{2 G M r}{c^2 a^2}
}\frac{\pi}{\sqrt{u_1^{\prime}}}F\left(\frac{1}{2},\frac{1}{2},1,
\frac{u_2^{\prime}}{u_1^{\prime}}\right) \nonumber \\
&+&4\frac{L}{2 \sqrt{Q} \sqrt{u_1^{\prime}}}\pi F_1\left(\frac{1}{2},1,
\frac{1}{2},1,\chi_1,\chi_2\right) \nonumber \\
\end{eqnarray}
where $\chi_1=\frac{1}{u_1^{\prime}},\;\chi_2=\frac{u_2^{\prime}}{u_1^{\prime}}$.

We have calculated the amount of frame dragging for galactic black holes
for the values of Kerr parameter given in equation (\ref{Galaxy}) and for 
fixed values 
for radii and Carter's 
constant $Q$. The invariant parameters $L,E$ are determined by the 
two conditions for spherical orbits. We repeated the analysis for 
different values of the Kerr parameter \cite{Porquet}. The results are 
displayed in table 4.

\begin{table}
\begin{center}
\begin{tabular}{|c|c|}\hline\hline
{\bf parameters} & {\bf predicted dragging}  \\

$a=0.52,L=-2.03566, E=0.957665, Q=11$ & $\Delta \phi^{{\rm GTR}}=-6.06933=
-1.2519\times10^6\frac{{\rm arcs}}{{\rm revolution}}$ \\
$a=0.9939, L=-2.25773, E=0.959284,Q=11$ 
& $\Delta \phi^{{\rm GTR}}=-5.86166=-1.2090\times 10^6\frac{{\rm arcs}}{{\rm revolution}}$ \\
$a=0.99616, L=-2.25883, E=0.959292,Q=11$ 
& $\Delta\phi^{\rm GTR}=-5.86161=-1.20904\times10^6\frac{{\rm arcs}}{{\rm revolution}} $  \\
\hline \hline
\end{tabular}
\end{center}
\caption{Predictions for frame dragging from a galactic black hole, with the 
indicated values of
Kerr parameter $a_{{\rm Galactic}}$, for  different 
values of test particle's angular momentum, a particular 
value for Carter's constant and 
for a fixed radius $r=10$. The values of the radii $a$,$L$ are in units of $GM_{{\rm BH}}/c^2$, 
while Carter's constant ${\cal Q}$ in units 
of $(GM_{{\rm BH}}/c^2)^2$.  The period ratios, $\tau$, are $3.058$i,$2.69743$i,
$2.69625$i respectively.}
\label{EINSTEIN1SNP}
\end{table}

\subsection{General solution for non-spherical geodesics in Kerr metric}

In the general case with non-zero 
cosmological constant, of non-spherical orbits one has to solve the differential equations
\begin{equation}
\int^\theta \frac{d\theta}{\sqrt{\Theta^{\prime}}}=\int^r \frac{dr}{\sqrt{R^{\prime}}}
\label{genulllike}
\end{equation}
where $R^{\prime}(r)$ is a quartic polynomial for null geodesics ($\mu=0$ in 
Eq.(\ref{LARTH}))
which is given by
\begin{equation}
R^{\prime}=E^2\left\{\Xi^2 \left[(r^2+a^2)-a \Phi\right]^2-\Delta_r 
\left[ \Xi^2(\Phi-a)^2+ {\cal Q}\right]\right\}
\end{equation}
and $\Theta^{\prime}(\theta)$ is given 
by 
\begin{equation}
\Theta^{\prime}=E^2\left\{\left[{\cal Q}+(\Phi-a)^2\Xi^2\right]\Delta_{\theta}
-\Xi^2\frac{(a\sin^2 \theta-\Phi)^2}{\sin^2\theta}\right\}
\end{equation}

Note that this equation is an equation 
between two elliptic integrals, in the general case when 
all roots are distinct. The left hand side can be transformed into 
an elliptic integral with variable $z$ or $\xi$ in Weierstra$\ss$ normal form, 
see Eq.(\ref{subwei}).
In order to solve this differential equation and determine $r$ as a function 
of $\theta$, one can employ the theory 
of Abel for the transformation of elliptic functions \cite{NHA}, which was 
first applied in \cite{KraniotisKerr} for the case of timelike orbits. A detailed exposition of the theory of transformation of elliptic functions based 
on \cite{NHA} can also be found in \cite{KraniotisKerr}. The interesting 
connection with modular equations \cite{KraniotisKerr} is outlined in 
Appendix  \ref{MODULAR}.

We note at this point that the corresponding relationship for non-spherical 
timelike orbits in the presence of the cosmological constant relates 
a hyperelliptic integral (the polynomial $R^{\prime}$ is a sectic 
in this case) to an elliptic integral, and therefore involves 
the transformation theory of hyperelliptic functions. This fact was 
first observed in \cite{KraniotisKerr} and will be a subject of another 
publication.

\subsection{Transforming the geodesic elliptic integrals into Abel's form}

The transformation
\begin{equation}
x\rightarrow e_3+\frac{(e_1-e_3)}{x^2}
\end{equation}
transforms the elliptic integral in Weierstra$\ss$ form into Abel's and Jacobi's form:
\begin{equation}
\int \frac{dx}{\sqrt{4(x-e_1)(x-e_2)(x-e_3)}}\rightarrow -\frac{1}{\sqrt{e_1-e_3}}\int \frac{dx}{\sqrt{(1-x^2)(1-k^2 x^2)}}
\end{equation}
with $k^2=\frac{e_2-e_3}{e_1-e_3}$.

Similarly, the quartic can be brought to Jacobi's form $y^2=(1-x^2)(1-k^2_1 x^2)$.



Indeed, as we saw in section 3 we have 
\begin{equation}
\frac{d\theta}{\sqrt{\Theta}}=-\frac{1}{2}\frac{dz}{\sqrt{z^3 \alpha-
z^2 (\alpha+\beta)+{\cal Q} z}}=-\frac{1}{2}\frac{d\xi}{\sqrt{4 \xi^3-g_2 \xi-
g_3}}
\end{equation}
where 
\begin{eqnarray}
g_2 &=& \frac{1}{12}\left(-a^2+{\cal Q}+\Phi^2\right)^2-\frac{{\cal Q}}{4}(-a^2) \nonumber \\
g_3 &=& \frac{1}{216} \left(-a^2+{\cal Q}+\Phi^2\right)^3
-\frac{{\cal Q}}{48} a^4-
\frac{{\cal Q}}{48} (-a^2 ({\cal Q}+\Phi^2)) \nonumber \\
\end{eqnarray}
In this case the three roots of the cubic $4 \xi^3-g_2 \xi-g_3$ are given 
by the expressions
\begin{eqnarray}
e_1 &=&\frac{1}{24} \left(-a^2+\Phi^2+Q+
3\sqrt{a^4+2 a^2 {\cal Q}+{\cal Q}^2-2 a^2 \Phi^2+2 
{\cal Q} \Phi^2+\Phi^4}\right) \nonumber \\
e_2 &=&\frac{1}{12}\left(a^2-{\cal  Q}-\Phi^2\right) \nonumber \\
e_3 &=& \frac{1}{24} \left(-a^2+\Phi^2+Q-
3\sqrt{a^4+2 a^2 {\cal Q}+{\cal Q}^2-2 a^2 \Phi^2+2 
{\cal Q} \Phi^2+\Phi^4}\right) \nonumber \\
\end{eqnarray}
Thus the Jacobi modulus $k^2=\frac{e_2-e_3}{e_1-e_3}$ is given by
\begin{equation}
k^2=\frac{a^2-{\cal Q}-\Phi^2+\sqrt{a^4+2 a^2 ({\cal Q}-\Phi^2)+({\cal Q}+
\Phi^2)^2}}{2\sqrt{a^4+2 a^2 ({\cal Q}-\Phi^2)+({\cal Q}+
\Phi^2)^2}}
\end{equation}
It has the correct limit for photon spherical geodesics with $\Phi=0$
\begin{equation}
k^2 (\Phi=0)=\frac{a^2 }{a^2+{\cal Q}}
\end{equation}

Also 
\begin{equation}
\frac{1}{\sqrt{e_1-e_3}}=\frac{2}{\left(a^4+2 a^2 ({\cal Q}-\Phi^2)+
({\cal Q}+\Phi^2)^2\right)^{1/4}}
\end{equation}
and it also has the correct limit for spherical photon orbits 
with zero $\Phi$ 
\begin{equation}
\frac{1}{\sqrt{e_1-e_3}}=\frac{2}{\sqrt{a^2+{\cal Q}}}
\end{equation}

Thus, we get 
\begin{equation}
\int \frac{d\theta}{\sqrt{\Theta}}=-\frac{1}{2}\frac{1}{\sqrt{e_1-e_3}}
\int \frac{d \xi^{\prime}}{\sqrt{(1-\xi^{\prime 2})(1-k^2 \xi^{\prime 2})}}
\end{equation}

Applying Luchterhand's 
transformation formula \cite{LUCH} on the radial integral the Jacobi's 
form can be recovered:
\begin{equation}
\frac{\partial x}{M \sqrt{(1-x^2)(1-k_1^2x^2)}}=\frac{\partial y}{
\sqrt{(y-\alpha)(y-\beta)(y-\gamma)(y-\delta)}}
\end{equation}
where the Jacobi modulus $k_1$ and the coefficient $M$, are given in terms of 
the roots $\alpha,\beta,\gamma,\delta$ of the quartic, by the following 
expressions
\begin{equation}
k_1=\frac{\sqrt{(\alpha-\delta)(\beta-\gamma)}}{\sqrt{(\alpha-\gamma)(
\beta-\delta)}},\;\;\;M=\sqrt{(\alpha-\gamma)(\beta-\delta)}
\end{equation}
and the integration variables are related by
\begin{equation}
\frac{1-x}{1+x}=\frac{(\gamma-\delta)(y-\alpha)(y-\beta)}{(\alpha-\beta)
(y-\gamma)(y-\delta)}
\end{equation}
Also we assume that the roots $\alpha,\beta,\gamma,\delta$ 
of the quartic are real and are organized in
the following ascending order of magnitude: $\alpha>\beta>\gamma>\delta$.

We can also provide a nice formula for $r$ in terms of the 
Jacobi's sinus amplitudinus function \footnote{We wright
$\int \frac{dr}{\sqrt{R}}=\int \frac{dr}{\sqrt{(r-\alpha)(r-\beta)(r-\gamma)(r-\delta)}}=\int \frac{\partial x}{M \sqrt{(1-x^2)(1-k_1^2 x^2)}}=
\int \frac{d\theta}{\sqrt{\Theta}}$.}
\begin{equation}
\frac{(\gamma-\delta)(r-\alpha)(r-\beta)}{(\alpha-\beta)(r-\gamma)
(r-\delta)}=\frac{1-{\rm sn}\left(M \int \frac{d\theta}{\sqrt{\Theta}},k_1\right)}{1+{\rm sn}\left(M\int \frac{d\theta}{\sqrt{\Theta}},k_1\right)}
\end{equation}

\subsection{Equatorial geodesics including the contribution of the
cosmological constant}

The equatorial geodesics (i.e. $\theta=\pi/2,Q=0$), with a 
nonzero cosmological constant,  may be obtained by  Eq.(\ref{LambdaGeo}) for 
the particular values of $Q,\theta$.
The  characteristic function in this case, has the form \cite{KraniotisKerr}

\begin{equation}
W=-Ect+\int\frac{\sqrt{R^{\prime}}}{\Delta_r}dr+L\phi
\end{equation}
and the geodesics are given by the expressions:

\begin{eqnarray}
\frac{dr}{\sqrt R^{\prime}}&=&\frac{d\lambda}{r^2}  \nonumber \\
r^2\frac{d\phi}{d\lambda} &=& \frac{a (1+\frac{1}{3}a^2\Lambda )^2(E(r^2
+a^2)-La) }{(1-\frac{\Lambda}{3}r^2)(r^2+a^2)-\frac{2 G M r}{c^2}}
+(L-a E)(1+\frac{1}{3}a^2 \Lambda )^2 \nonumber \\
c r^2\frac{dt}{d\lambda} &=&\frac{(1+\frac{1}{3}a^2\Lambda
)^2(r^2+a^2)\left[(r^2+a^2)E-a L\right]}{\Delta_r}
+(1+\frac{1}{3} a^2 \Lambda)^2 a (L-a E)  \nonumber \\
\label{equilambda}
\end{eqnarray}
where
\begin{equation}
R^{\prime}=(1+\frac{1}{3}a^2\Lambda )^2 \left[((r^2+a^2)E-a L)^2-
\Delta_r((L-a E)^2)\right]
\end{equation}
for null geodesics and 
\begin{equation}
R^{\prime}=(1+\frac{1}{3}a^2\Lambda )^2 \left[((r^2+a^2)E-a L)^2-
\Delta_r((L-a E)^2)\right]-\Delta_r(\mu^2r^2)
\end{equation}
for timelike geodesics.

Equatorial orbits are of particular interest for various astrophysics 
applications. The exact solution of circular equatorial orbits with 
 a cosmological constant was presented in \cite{KraniotisKerr}. 
In what follows, we shall concentrate on the cases of non-circular equatorial 
timelike and null orbits that describe motion of test particles and photons 
under the assumption of a vanishing cosmological constant. We shall derive 
new exact expressions for the precession of perihelion or periapsis 
for the 
orbit of test particle in the gravitational field of Kerr, as well as for 
the deflection angle of a light ray from the Kerr gravitational field, 
in terms of multivariable hypergeometric functions. In 
the latter case, we apply the exact calculation obtained for determining 
the bending angle of a light ray from the gravitational field of the galactic 
centre of Milky-Way assuming that the Sgr A$^{*}$ region is 
a supermassive rotating black hole for various values of the 
Kerr parameter which are supported 
by recent observations and of the impact factor.  
The more general case, in the presence of the cosmological constant is a task 
for a future publication.

\subsection{Exact solution of timelike equatorial geodesics}
\label{TIMEEQUATOR}

We now proceed to determine the exact expression for the precession 
of equatorial timelike orbits in Kerr spacetime.
We have $r^2 (\dot{r})=\sqrt{R}$. This can be rewritten as
\begin{equation}
(\dot{r})^2=E^2+\frac{a^2 E^2}{r^2}-\frac{L^2}{r^2}+\frac{2 GM}{c^2 r^3} 
(L-a E)^2-\frac{\Delta}{r^2}
\end{equation}
By defining a new variable $u=1/r$ we get the following expression
\begin{equation}
u^{-4}\dot{u}^2=E^2+a^2 E^2 u^2-L^2 u^2+\frac{2 GM}{c^2}(L-a E)^2 u^3
-(1+a^2 u^2-\frac{2 GM}{c^2}u)\equiv B^t(u)
\end{equation}
Similarly $\dot{\phi}^2=u^4 \frac{A^2(u)}{D^2(u)}$ where 
\begin{equation}
A(u)=L+u\alpha_S(a E-L),\;\;D(u)=1+a^2 u^2-\alpha_Su,\;\;\alpha_S:=
\frac{2 GM}{c^2}
\end{equation}
Thus, we obtain the differential equation
\begin{equation}
\frac{d\phi}{du}=\frac{A(u)}{D(u)}\frac{1}{\sqrt{B^t(u)}}
\end{equation}
We now write 
\begin{equation}
\frac{A(u)}{D(u)}=\frac{A_{+}}{u_{+}-u}+\frac{A_{-}}{u_{-}-u}
\end{equation}
where $u_{+}=\frac{r_{+}}{a^2},u_{-}=\frac{r_{-}}{a^2}$ and 
\begin{equation}
r_{\pm}=\frac{GM}{c^2}\pm \sqrt{\left(\frac{GM}{c^2}\right)^2-a^2}
\end{equation}
where the quantities $A_{+},A_{-}$ are given by
\begin{eqnarray}
A_{+}&=&\frac{\frac{L}{a^2}+\frac{\alpha_S}{a^2}\left(aE-L\right)u_{+}}
{u_{-}-u_{+}}
\nonumber \\
A_{-}&=&\frac{-\frac{L}{a^2}-\frac{\alpha_S}{a^2}\left(aE-L\right)u_{-}}
{u_{-}-u_{+}} \nonumber \\
\end{eqnarray}
For the calculation of the perihelion (periapsis) precession of a test particle in  orbit around a rotating mass we need to calculate the integral 
$\Delta \phi^{\rm GTR}=2\int_{u_3^{\prime}}^{u_2^{\prime}}d\phi$.
Then 
\begin{eqnarray}
\int d\phi&=&\int \frac{du^{\prime} A_{+}}{\left(\frac{Mr_{+}}{a^2}-u^{\prime}
\right)}\frac{1}{\sqrt{\frac{\alpha_S (L-a E)^2}{\left(\frac{GM}{c^2}\right)^3
}}}\frac{1}{\sqrt{(u^{\prime}-u_3^{\prime})(u_1^{\prime}-u^{\prime})
(u_2^{\prime}-u^{\prime})}} \nonumber \\
&+&\int \frac{du^{\prime} A_{-}}{\left(\frac{Mr_{-}}{a^2}-u^{\prime}
\right)}\frac{1}{\sqrt{\frac{\alpha_S (L-a E)^2}{\left(\frac{GM}{c^2}\right)^3
}}}\frac{1}{\sqrt{(u^{\prime}-u_3^{\prime})(u_1^{\prime}-u^{\prime})
(u_2^{\prime}-u^{\prime})}} \nonumber \\
\end{eqnarray}
and we have defined $u^{\prime}:=u\frac{GM}{c^2}$.
Using new variables
$$u^{\prime}\rightarrow u_3^{\prime}+\xi^2 (u_2^{\prime}-u_3^{\prime})$$
we can bring our expression to the integral representation of Appell's first 
hypergeometric function
\begin{eqnarray}
\Delta\phi^{\rm GTR}&=&\frac{2}{\sqrt{u_1^{\prime}-u_3^{\prime}}}
\frac{1}{\sqrt{\frac{\alpha_S (L-a E)^2}{\left(\frac{GM}{c^2}\right)^3}}}
\Biggl\{\frac{A_{+}}{\frac{GM r_{+}}{c^2 a^2}-u_3^{\prime}}
F_1\left(\frac{1}{2},1,\frac{1}{2},1,\frac{u_2^{\prime}-u_3^{\prime}}
{\frac{GM r_{+}}{c^2 a^2}-u_3^{\prime}},\frac{u_2^{\prime}-u_3^{\prime}}{
u_1^{\prime}-u_3^{\prime}}\right)\frac{\Gamma(1/2)^2}{\Gamma(1)}
\nonumber \\
&+&\frac{A_{-}}{\frac{GM r_{-}}{c^2 a^2}-u_3^{\prime}}
F_1\left(\frac{1}{2},1,\frac{1}{2},1,\frac{u_2^{\prime}-u_3^{\prime}}
{\frac{GM r_{-}}{c^2 a^2}-u_3^{\prime}},\frac{u_2^{\prime}-u_3^{\prime}}{
u_1^{\prime}-u_3^{\prime}}\right)\frac{\Gamma(1/2)^2}{\Gamma(1)}\Biggr\}
\nonumber \\
\end{eqnarray}

We can now provide an exact expression for time $t$. The first 
differential equation in Eq.(\ref{kerrgeo}) for equatorial orbits 
can be written in terms of the variable $u$ as follows
\begin{equation}
c\dot{t}=\frac{E(1+a^2 u^2)+a \alpha_S u^3 (a E-L)}{D{u}}
\end{equation}
Subsequently by dividing with $\dot{u}$ we derive the equation
\begin{eqnarray}
\frac{cdt}{du}&=&\frac{E(1+a^2 u^2)+a \alpha_S u^3 (a E-L)}{
u^2 D(u)\sqrt{B^t (u)}} \nonumber \\
&=&\frac{E}{u^2 D(u)\sqrt{B^t (u)}}+\frac{E a^2}{D(u)\sqrt{B^t (u)}}+
\frac{a u\alpha_S (a E-L)}{D(u)\sqrt{B^t (u)}} \nonumber \\
\end{eqnarray}
By integrating we can express $t$ in terms of Appell's and Lauricella 
generalized hypergeometric functions

\begin{eqnarray}
ct&=&\frac{2}{\sqrt{u_1^{\prime}-u_3^{\prime}}}
\frac{1}{\sqrt{\frac{\alpha_S (L-a E)^2}{\left(\frac{GM}{c^2}\right)^3}}}
\Biggl\{\frac{A_{+}^t}{\frac{GM r_{+}}{c^2 a^2}-u_3^{\prime}}
F_1\left(\frac{1}{2},1,\frac{1}{2},1,\frac{u_2^{\prime}-u_3^{\prime}}
{\frac{GM r_{+}}{c^2 a^2}-u_3^{\prime}},\frac{u_2^{\prime}-u_3^{\prime}}{
u_1^{\prime}-u_3^{\prime}}\right)\pi
\nonumber \\
&+&\frac{A_{-}^t}{\frac{GM r_{-}}{c^2 a^2}-u_3^{\prime}}
F_1\left(\frac{1}{2},1,\frac{1}{2},1,\frac{u_2^{\prime}-u_3^{\prime}}
{\frac{GM r_{-}}{c^2 a^2}-u_3^{\prime}},\frac{u_2^{\prime}-u_3^{\prime}}{
u_1^{\prime}-u_3^{\prime}}\right)\pi\Biggr\}
\nonumber \\
&+&2\frac{A_{+}^{\prime}}{\sqrt{u_1^{\prime}-u_2^{\prime}}}
\frac{1}{\frac{u_2^{\prime 2}}{G^2 M^2 c^{-4}}}\frac{1}
{\sqrt{\frac{\alpha_S (L-a E)^2}{\left(\frac{GM}{c^2}\right)^3}}}
\Biggl\{\frac{1}{\frac{G M r_{+}}{c^2 a^2}-u_2^{\prime}}
F_D\left(\frac{1}{2},2,1,\frac{1}{2},1,\frac{u_2^{\prime}-u_3^{\prime}}{
u_2^{\prime}},\frac{u_3^{\prime}-u_2^{\prime}}{\frac{G M r_{+}}{c^2 a^2}-u_2^{\prime}},\frac{u_3^{\prime}-u_2^{\prime}}{u_1^{\prime}-u_2^{\prime}}
\right)\pi \nonumber \\
&-&\frac{1}{\frac{G M r_{-}}{c^2 a^2}-u_2^{\prime}}
F_D\left(\frac{1}{2},2,1,\frac{1}{2},1,\frac{u_2^{\prime}-u_3^{\prime}}{
u_2^{\prime}},\frac{u_3^{\prime}-u_2^{\prime}}{\frac{G M r_{-}}{c^2 a^2}-u_2^{\prime}},\frac{u_3^{\prime}-u_2^{\prime}}{u_1^{\prime}-u_2^{\prime}}
\right)\pi\Biggr\}
\end{eqnarray}

\subsection{Exact solution of null equatorial geodesics}
\label{PHOSEQUATOR}

In this case, we arrive at the differential equation first derived in
\cite{Boyer}
\begin{equation}
\frac{d\phi}{du}=\frac{\Phi+u \alpha_S (a-\Phi)}{D(u)}\frac{1}{\sqrt{B^N(u)}}
\label{NullEquator}
\end{equation}
where the cubic polynomial $B^N(u)$ is given by the expression
\begin{equation}
B^N(u)=\alpha_S (\Phi-a)^2 u^3+u^2 (a^2-\Phi^2)+1
\label{BNULL}
\end{equation}
In order to calculate the angle of deflection is  necessary to 
calculate the integral:
$\Delta\phi^{{\rm GTR}}=2 \int_0^{u_2^{\prime}}d\phi$.

Using partial fractions as in the timelike case we obtain an elegant 
exact expression for $\Delta\phi^{\rm GTR}$ in terms of Lauricella's fourth, 
hypergeometric function of three variables $F_D$, and Appell's first 
hypergeometric function of two variables $F_1$,

\begin{eqnarray}
\Delta \phi^{{\rm GTR}}&=&\frac{2}{\sqrt{u_1^{\prime}-u_3^{\prime}}}
\frac{1}{\sqrt{\frac{\alpha_S(\Phi-a)^2}{\frac{G^3 M^3}{c^6}}}}
\Biggl(\frac{A_{+}}{\frac{G M r_{+}}{c^2 a^2}-u_3^{\prime}}
F_1\left(\frac{1}{2},1,\frac{1}{2},1,\frac{u_2^{\prime}-u_3^{\prime}}
{\frac{G M r_{+}}{c^2 a^2}-u_3^{\prime}},\frac{u_2^{\prime}-u_3^{\prime}}
{u_1^{\prime}-u_3^{\prime}}\right)\frac{\Gamma\left(\frac{1}{2}\right)
\Gamma\left(\frac{1}{2}\right)}{\Gamma(1)} \nonumber \\
&+&\frac{A_{-}}{\frac{G M r_{-}}{c^2 a^2}-u_3^{\prime}}
F_1\left(\frac{1}{2},1,\frac{1}{2},1,\frac{u_2^{\prime}-u_3^{\prime}}
{\frac{G M r_{-}}{c^2 a^2}-u_3^{\prime}},\frac{u_2^{\prime}-u_3^{\prime}}
{u_1^{\prime}-u_3^{\prime}}\right)\frac{\Gamma \left(\frac{1}{2}\right)
\Gamma \left(\frac{1}{2}\right)}{\Gamma(1)}\Biggr) \nonumber \\
&+&\frac{2}{\sqrt{u_1^{\prime}-u_3^{\prime}}}
\frac{1}{\sqrt{\frac{\alpha_S(\Phi-a)^2}{\frac{G^3 M^3}{c^6}}}} \nonumber \\
&\Biggl(&-A_{+}\sqrt{\frac{-u_3^{\prime}}{u_2^{\prime}-u_3^{\prime}}} 
\frac{1}{\left(\frac{G M r_{+}}{c^2 a^2}-u_3^{\prime}\right)}2 
F_D\left(\frac{1}{2},1,\frac{1}{2},\frac{1}{2},\frac{3}{2},
\frac{-u_3^{\prime}}{\frac{G M r_{+}}{c^2 a^2}-u_3^{\prime}},
\frac{-u_3^{\prime}}{u_1^{\prime}-u_3^{\prime}},
\frac{-u_3^{\prime}}{u_2^{\prime}-u_3^{\prime}}\right) \nonumber \\
&-&A_{-} \sqrt{\frac{-u_3^{\prime}}{u_2^{\prime}-u_3^{\prime}}}
\frac{1}{\left(\frac{G M r_{-}}{c^2 a^2}-u_3^{\prime}\right)}2 
F_D\left(\frac{1}{2},1,\frac{1}{2},\frac{1}{2},\frac{3}{2},
\frac{-u_3^{\prime}}{\frac{G M r_{-}}{c^2 a^2}-u_3^{\prime}},
\frac{-u_3^{\prime}}{u_1^{\prime}-u_3^{\prime}},
\frac{-u_3^{\prime}}{u_2^{\prime}-u_3^{\prime}}\right)\Biggr) \nonumber \\
\end{eqnarray}
or 

\begin{eqnarray}
\Delta \phi^{{\rm GTR}}&=&\frac{2}{\sqrt{u_1^{\prime}-u_3^{\prime}}}
\frac{1}{\sqrt{\frac{\alpha_S(\Phi-a)^2}{\frac{G^3 M^3}{c^6}}}}
\Biggl\{\frac{A_{+}}{\frac{G M r_{+}}{c^2 a^2}-u_3^{\prime}}
\Biggl(F_1\left(\frac{1}{2},1,\frac{1}{2},1,\frac{u_2^{\prime}-u_3^{\prime}}
{\frac{G M r_{+}}{c^2 a^2}-u_3^{\prime}},\frac{u_2^{\prime}-u_3^{\prime}}
{u_1^{\prime}-u_3^{\prime}}\right)\pi \nonumber \\
&-&2  \sqrt{\frac{-u_3^{\prime}}{u_2^{\prime}-u_3^{\prime}}}
F_D\left(\frac{1}{2},1,\frac{1}{2},\frac{1}{2},\frac{3}{2},
\frac{-u_3^{\prime}}{\frac{G M r_{+}}{c^2 a^2}-u_3^{\prime}},
\frac{-u_3^{\prime}}{u_1^{\prime}-u_3^{\prime}},
\frac{-u_3^{\prime}}{u_2^{\prime}-u_3^{\prime}}\right) \Biggr) \nonumber \\
&+&\frac{A_{-}}{\frac{G M r_{-}}{c^2 a^2}-u_3^{\prime}}
\Biggl(F_1\left(\frac{1}{2},1,\frac{1}{2},1,\frac{u_2^{\prime}-u_3^{\prime}}
{\frac{G M r_{-}}{c^2 a^2}-u_3^{\prime}},\frac{u_2^{\prime}-u_3^{\prime}}
{u_1^{\prime}-u_3^{\prime}}\right)\pi \nonumber \\
&-&2  \sqrt{\frac{-u_3^{\prime}}{u_2^{\prime}-u_3^{\prime}}}
F_D\left(\frac{1}{2},1,\frac{1}{2},\frac{1}{2},\frac{3}{2},
\frac{-u_3^{\prime}}{\frac{G M r_{-}}{c^2 a^2}-u_3^{\prime}},
\frac{-u_3^{\prime}}{u_1^{\prime}-u_3^{\prime}},
\frac{-u_3^{\prime}}{u_2^{\prime}-u_3^{\prime}}\right) \Biggr)\Biggr\} \nonumber \\
\label{DeflectionKerr}
\end{eqnarray}
During the derivation of Eq.(\ref{DeflectionKerr}) we have used at an intermediate 
step of the calculation the identity
\begin{eqnarray}
&&\frac{1}{\frac{Mr_{\pm}}{a^2}\sqrt{u_1^{\prime}u_2^{\prime}}}
F_D\left(1,1,\frac{1}{2},\frac{1}{2},\frac{3}{2},\frac{u_3^{\prime}}{
\frac{Mr_{\pm}}{a^2}},\frac{u_3^{\prime}}{u_1^{\prime}},\frac{u_3^{\prime}}
{u_2^{\prime}}\right) \nonumber \\
&=&\frac{1}{\frac{Mr_{\pm}}{a^2}-u_3^{\prime}\sqrt{u_1^{\prime}-
u_3^{\prime}}\sqrt{u_2^{\prime}-u_3^{\prime}}}
F_D\left(\frac{1}{2},1,\frac{1}{2},\frac{1}{2},\frac{3}{2},
\frac{-u_3^{\prime}}{\frac{Mr_{\pm}}{a^2}-u_3^{\prime}},
\frac{-u_3^{\prime}}{u_1^{\prime}-u_3^{\prime}},
\frac{-u_3^{\prime}}{u_2^{\prime}-u_3^{\prime}}\right) \nonumber \\
\end{eqnarray}

The quantities $A_{\pm}$ are now given in terms of the impact parameter 
$\Phi$ and the Kerr parameter $a$ by the expressions
\begin{equation}
A_{\pm}=\frac{\pm \Phi\pm\alpha_S(a-\Phi)\frac{r_{\pm}}{a^2}}{
-2 \sqrt{\left(\frac{GM}{c^2}\right)^2-a^2}}
\end{equation}

The angle of deflection $\delta$ of light-rays 
from the gravitational field of a Galactic rotating black hole or 
a massive star is defined to be the deviation of $\Delta \phi^{\rm GTR}$ 
from the transcendental number $\pi$
\begin{equation}
\delta=\Delta\phi^{\rm GTR}-\pi
\label{Deflection}
\end{equation}

We have calculated the deflection angle $\delta$ of light 
rays from the gravitational field of a galactic rotating black hole 
for different values 
of the Kerr parameter $a$ and the impact parameter $\Phi$.
The results are displayed in figure (\ref{Deflection}) and tables \ref{EINSTEINDefle1},\ref{EINSTEINDefle2}. It is clear from  figure (\ref{Deflection}) that 
especially for smaller values of the impact parameter $\Phi$, there is a
strong dependence of the deflection angle on the spin of the black hole. 
This has implications for gravitational lensing studies and can lead in 
principle to an 
independent measurement of the Kerr parameter at the strong field regime.

\begin{table}
\begin{center}
\begin{tabular}{|c|c|}\hline\hline
{\bf parameters} &   {\bf predicted deflection} \\
$a_{{\rm Galactic}}=0.52, \Phi=5$ & 
$\Delta \phi^{{\rm GTR}}-\pi=1.84869=105.9^{\circ}=381319
{\rm arcs}$ \\
$a_{{\rm Galactic}}=0.52, \Phi=10$ &$\Delta \phi^{{\rm GTR}}-\pi=0.537=
30.77^{\circ}=110792
{\rm arcs}$ \\
$a_{{\rm Galactic}}=0.52, \Phi=15$ & $\Delta \phi^{{\rm GTR}}-\pi=0.320=
18.34^{\circ}=66015.6
{\rm arcs}$ \\
$a_{{\rm Galactic}}=0.52, \Phi=20$ & $\Delta \phi^{{\rm GTR}}-\pi=0.228=
13.08^{\circ}=47093{\rm arcs}$ \\
$a_{{\rm Galactic}}=0.52, \Phi=40$ & $\Delta \phi^{{\rm GTR}}-\pi=0.1065=
6.10^{\circ}=21973
{\rm arcs}$ \\
\hline \hline
\end{tabular}
\end{center}
\caption{Predictions for light deflectiom from a galactic rotating 
black hole with Kerr parameter $a_{{\rm Galactic}}=0.52\frac{GM_{{\rm BH}}}{c^2}$.
The values of the impact parameter $\Phi$ are in units of $\frac{GM_{{\rm BH}}}{c^2}$.}
\label{EINSTEINDefle1}
\end{table}

\begin{table}
\begin{center}
\begin{tabular}{|c|c|}\hline\hline
{\bf parameters} &   {\bf predicted deflection} \\
$a_{{\rm Galactic}}=0.9939, \Phi=5$ & 
$\Delta \phi^{{\rm GTR}}-\pi=1.317=75.47^{\circ}=271710
{\rm arcs}$ \\
$a_{{\rm Galactic}}=0.9939, \Phi=10$ &$\Delta \phi^{{\rm GTR}}-\pi=0.497=
28.49^{\circ}=102581
{\rm arcs}$ \\
$a_{{\rm Galactic}}=0.9939, \Phi=15$ & $\Delta \phi^{{\rm GTR}}-\pi=0.306=
17.57^{\circ}=63254
{\rm arcs}$ \\
$a_{{\rm Galactic}}=0.9939, \Phi=20$ & $\Delta \phi^{{\rm GTR}}-\pi=0.2217=
12.70^{\circ}=45728{\rm arcs}$ \\
$a_{{\rm Galactic}}=0.9939, \Phi=40$ & $\Delta \phi^{{\rm GTR}}-\pi=0.1052=
6.026^{\circ}=21694.3
{\rm arcs}$ \\
\hline \hline
\end{tabular}
\end{center}
\caption{Predictions for light deflectiom from a galactic rotating 
black hole with Kerr parameter $a_{{\rm Galactic}}=0.9939\frac{GM_{{\rm BH}}}{c^2}$.
The values of the impact parameter $\Phi$ are in units of $\frac{GM_{{\rm BH}}}{c^2}$.}
\label{EINSTEINDefle2}
\end{table}

\begin{figure}
\epsfxsize=6.0in
\epsfysize=6.5in
\epsffile{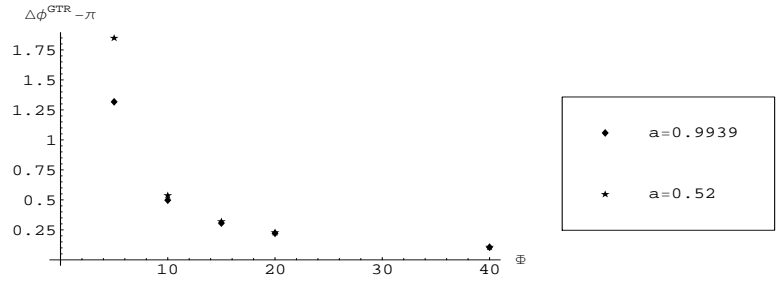}
\caption{Deflection angle $\delta$ for various values of impact parameter 
$\Phi$ and for two values of the Kerr parameter $a$ for the rotating 
black hole.}
\label{Deflection}
\end{figure}


The roots of the cubic are organized as $u_1^{\prime}>u_2^{\prime}>0>u_3^{\prime}$.

\subsubsection{Null equatorial geodesics with a cosmological constant}

In this case, the generalization of equation (\ref{NullEquator}) is given 
by
\begin{equation}
\frac{d\phi}{du}=\frac{\Phi u^2-\alpha_S u^3(\Phi-a)+(\Phi-a)(-\frac{\Lambda}{3})(1+a^2 u^2)}{
\left[\left(u^2-\frac{\Lambda}{3}\right)(1+a^2u^2)-\alpha_S u^3\right]\sqrt{B^{\Lambda}(u)}}
\end{equation}
where $B^{\Lambda}(u)$ is still a cubic polynomial 
\begin{equation}
B^{\Lambda}(u)=1+a^2 u^2-\Phi^2 u^2+\alpha_S u^3 (\Phi-a)^2+
\frac{\Lambda}{3}(1+a^2 u^2)(\Phi-a)^2
\end{equation}
For a vanishing cosmological constant the above cubic polynomial reduces to 
Eq.(\ref{BNULL}).

\section{Conclusions}
\label{SYMBE}

In this work, we have investigated the motion of a 
test particle and 
light in the gravitational field of Kerr spacetime with and without the cosmological 
constant. We have derived a number of useful analytical expressions for measurable physical quantities.

In the case of null orbits, we solved exactly the geodesic equations 
for spherical polar and non-polar photon orbits.The exact solution for 
the orbit of 
a photon with zero angular momentum 
$\Phi$ and vanishing cosmological constant 
was provided by the Weierstra$\ss$ elliptic function $\wp(z)$.

The exact 
expressions that determine the  amount of frame-dragging (Lense-Thirring 
effect) 
for the corresponding photon orbits, 
assuming vanishing cosmological constant, were 
written in 
terms of Weierstra$\ss$ function real half-period or equivalently in terms of 
Gau$\ss$ hypergeometric function $F$ for a 
photon spherical orbit with $\Phi=0$ and  Gau$\ss$ hypergeometric 
function $F$ and 
Appell's generalized hypergeometric 
function of two variables $F_1$ for photon orbits with $\Phi\not =0$ and 
constant radius.
The corresponding expressions in the presence of the cosmological constant 
were given in terms of Appell's hypergeometric function $F_1$ for the 
case of spherical null polar orbits, and in terms of Lauricella's 
hypergeometric function of three variables  $F_D$, Appell's 
$F_1$ and Gau$\ss$ ordinary hypergeometric function $F$, in the case 
of spherical photonic orbits with nonvanishing value for the invariant 
parameter $\Phi$. 

We subsequently applied our  exact solutions for the determination of the Lense-Thirring effect that a photon experiences in a spherical polar and non-polar orbit
around and close to our galactic centre, assuming the latter is a rotating black hole 
whose surrounding spacetime structure is described by the Kerr geometry as supported 
by recent observations. We repeated the analysis for various values of the 
Kerr parameter.

We also solved exactly non-spherical polar null unbound orbits. We derived 
analytical results for the deflection angle of a light ray from the 
gravitational field of a rotating black hole's pole. The resulting expression 
for the deflection angle was written elegantly in terms of Lauricella's 
hypergeometric function $F_D$.

We then investigated, non-circular orbits confined to the equatorial plane 
(timelike and null-like equatorial geodesics) for which the 
value of Carter's constant invariant parameter vanishes. For the case
of a vanishing cosmological constant, we derived an exact expression
for the amount of relativistic precession for a test particle in a 
timelike orbit, around a  rotating 
central mass.  The corresponding novel expression was  given 
in terms of Appell's first hypergeometric function of two variables 
$F_1$.
The application of this exact solution as well as of those 
that describe non-spherical orbits not necessarily confined 
to the equatorial plane, for the determination 
of the effect of rotation of central mass on the perihelion precession 
of a test particle (Mercury around Sun) or periapsis precession 
for a star such as S2 in a high eccentricity 
orbit around the galactic centre \cite{Schodel} is 
beyond the scope of this work and will be the subject of a future 
publication \cite{KRANIOTIS}.

We have also derived an exact expression for the deflection angle 
of  a light ray from   the gravitational 
field of a rotating mass (the Kerr field).
The corresponding expression was given in terms of Lauricella's $F_D$ 
and Appell's $F_1$ generalized hypergeometric functions.
 We applied this calculation for 
the bending of light  from the gravitational field of the galactic centre 
of Milky Way, assuming the latter is a supermassive Kerr black hole 
for various values of the Kerr parameter and the impact factor.
We emphasized in the main text the strong dependence of the bending 
angle on the Kerr parameter for small values of the impact factor.
These results should be useful for gravitational lensing studies,
where one treats the black hole as a gravitational ``lens'' \cite{Ohanian}, 
especially in the strong field region, when the bending 
angle can be very large.

The synergy between theory and experiment for probing and measuring relativistic  effects is going to be one of the most exciting and fruitful scientific 
endeavours.

\section*{Acknowledgements} 
This work is supported by DOE grant DE-FG03-95-Er-40917.

\appendix

\section{Definitions of genus-2 theta functions that solve Jacobi's inversion 
problem}

Riemann's theta function \cite{BerGeorgR} for genus $g$ is defined as follows:

\begin{equation}
\Theta(u):=\sum_{n_1,\cdots,n_g} e^{2\pi i u n+i \pi \Omega n^2}
\end{equation}
where $\Omega n^2:=\Omega_{11} n_1^2+\cdots 2 \Omega_{12}n_1 n_2 +\cdots$ 
 and $un:=u_1n_1+\cdots u_g n_g$. The symmetric $g\times g$ complex matrix 
$\Omega$ whose imaginary part is positive definite is a member of the 
set called Siegel upper-half-space denoted as ${\cal L}_{S_g}$. It is 
clearly the generalization of the ratio of half-periods $\tau$ in the genus $g=1$ case.    
For genus $g=2$ the Riemann theta function can be written in 
matrix form:

\begin{eqnarray}
\Theta(u,\Omega)&=&\sum_{{\bf{n}} \in Z^2} e^{\pi i {\bf ^{t} n}\Omega {\bf n}+
2\pi i {\bf ^{t} n} {\bf u}} \nonumber \\
&=& \sum_{n_1,n_2} e^{\pi i \left(\begin{array}{cc}n_1 & n_2\end{array}\right) \left(\begin{array}{cc}
\Omega_{11} & \Omega_{12} \\
\Omega_{12} &  \Omega_{22} \end{array}\right) \left(\begin{array}{c}
n_1 \\
n_2\end{array}\right) +2 \pi i \left(\begin{array}{cc} n_1 &n_2 \end{array}\right) \left(\begin{array}{c}
u_1 \\
u_2\end{array}\right)} \nonumber \\
\end{eqnarray}
Riemann's theta function with characteristics is defined by:
\begin{equation}
\Theta(u;q,q^{\prime}):=\sum_{n_1,\cdots,n_g}e^{2\pi i u(n+q^{\prime})+
i\pi \Omega (n+q^{\prime})^2+2 \pi i q(n+q^{\prime})}
\label{thechara}
\end{equation}
herein $q$ denotes the set of $g$ quantities $q_1,\cdots,q_g$ and $q^{\prime}$ 
denotes the set of $g$ quantities $q^{\prime}_{1},\cdots,q^{\prime}_g$.
Eq.(\ref{thechara}) can be rewritten in a suggestive matrix form:
\begin{equation}
\Theta \left [\begin{array}{c} q^{\prime} \\ q \end{array} \right](u,\Omega)=
\sum_{n \in Z^g} e^{\pi i ^{t} (n+q^{\prime}) \Omega (n+q^{\prime})+
2 \pi i ^{t} (n+q^{\prime})(u+q)}, \;\;\;\;q,q^{\prime} \in Q^g
\end{equation}

The theta functions whose quotients provide a solution to Abel-Jacobi's inversion 
problem are defined as follows \cite{BAKER}:
\begin{equation}
\theta(u;q,q^{\prime}):=\sum e^{au^2+2 hu(n+q^{\prime})+b(n+q^{\prime})^2+
2 i\pi q(n+q^{\prime})}
\end{equation}
where the summation extends to all positive and negative integer values of 
the $g$ integers $n_1,\cdots,n_g$, $a$ is any symmetrical matrix whatever 
of $g$ rows and columns, $h$ is any matrix whatever of $g$ rows and columns, 
in general not symmetrical, $b$ is any symmetrical matrix whatever of $g$ 
rows and columns, such that the real part of the quadratic form  $bm^2$ 
is necessarily negative for all real values of the quantities $m_1,\cdots,
m_g$, other than zero, and $q,q^{\prime}$ constitute the characteristics of 
the function. The matrix $b$ depends on $\frac{1}{2}g (g+1)$ independent 
constants; if we put $i\pi \Omega=b$ and denote the $g$-quantities 
$hu$ by $i\pi U$, we obtain the relation with Riemann's theta function:
\begin{equation}
\theta(u;q,q^{\prime})=e^{au^2}\Theta(U;q,q^{\prime})
\end{equation}

The dependence of genus-2 theta functions on two complex variables is
denoted by:  $\theta(u;q,q^{\prime})=\theta(u_1,u_2;q,q^{\prime})$,
the dependence on the Siegel moduli matrix $\Omega$ by:
$\theta(u_1,u_2,\Omega;q,q^{\prime})$.
To every half-period one can associate a set of characteristics.
For instance, the period $u^{a,a_1}=\frac{1}{2}\left(\begin{array}{cc}
1 & 0 \\
1 & 0 \end{array}\right)$ while $\theta(u)$ is a theta function of 
two variables with  zero
characteristics, i.e. $\theta(u)=\theta(u;0,0)=\theta \left
[\begin{array}{cc} 0 & 0 \\ 0 & 0 \end{array} \right](u,\Omega)$.
Also,
Weierstra$\ss$ had associated a symbol for each of the six odd theta
functions with characteristics 
and the ten even theta functions of genus two. For example,
$\theta(u)$ is associated with the Weierstra$\ss$ symbol 5 or
occasionaly the number appears as a subscript, i.e. $\theta(u)_5$.

Let the genus $g$  Riemann hyperelliptic surface be described by the 
equation:
\begin{equation}
y^2=4 (x-a_1)\cdots (x-a_g) (x-c) (x-c_1)\cdots (x-c_g)
\label{Riemann}
\end{equation}
For $g=2$ the above hyperelliptic Riemann algebraic equation reduces to:
\begin{equation}
y^2=4 (x-a_1) (x-a_2) (x-c) (x-c_1) (x-c_2)
\label{Riemann}
\end{equation}
where $a_1,a_2,c_,c_1,c_2$ denote the finite branch points of the surface.

The Jacobi's inversion problem involves  finding the  solutions,
for $x_i$ in terms of $u_i$,  
for the following system of equations of Abelian integrals \cite{BAKER}:
\begin{eqnarray}
u_1^{x_1,a_1} & +&\cdots + u_1^{x_g,a_g} \equiv  u_1 \nonumber \\
\vdots        &+ &\cdots +  \vdots \;\;\;\;\;\;\;\;\;\;\;\;\;  \vdots \nonumber \\
u_g^{x_1,a_1}  &+&\cdots + u_g^{x_g,a_g} \equiv u_g
\end{eqnarray}
where 
$u_1^{x,\mu}=\int_{\mu}^{x}\frac{dx}{y},
u_2^{x,\mu}=\int_{\mu}^{x}\frac{x dx}{y},
\cdots,u_g^{x,\mu}=\int_{\mu}^{x}\frac{x^{g-1} dx}{y}$.

For $g=2$ the above system of equations takes the form:
\begin{eqnarray}
\int_{a_1}^{x_1} \frac{dx}{y}+\int_{a_2}^{x_2} \frac{dx}{y}\equiv u_1 \nonumber \\
\int_{a_1}^{x_1} \frac{x\;dx}{y}+\int_{a_2}^{x_2} \frac{x\;dx}{y} \equiv u_2
\label{Umkehr}
\end{eqnarray}
where $u_1,u_2$ are arbitrary.
The solution  is given by the five equations \cite{BAKER} 
\begin{eqnarray}
\frac{ \theta^2(u|u^{b,a})}{
\theta^2(u)}&=&A(b-x_1)(b-x_2)\cdots(b-x_g) \nonumber \\
&=&A(b-x_1)(b-x_2) \nonumber \\
&=&\pm \frac{(b-x_1)(b-x_2)}{\sqrt{e^{\pi i P P^{\prime}}f^{\prime}(b)}};
\label{inveb}
\end{eqnarray}
where $f(x)=(x-a_1)(x-a_2)(x-c)(x-c_1)(x-c_2)$,
and $e^{\pi i P P^{\prime}}=\pm 1$ accordingly as $u^{b,a}$ is an odd or even 
half-period. Also $b$ denotes a finite branch point and the branch
place $a$ being at infinity \cite{BAKER}.
The symbol $\theta(u|u^{b,a})$ denotes a genus 2 theta function with
characteristics: $\theta(u;q,q^{\prime})$ \cite{BAKER}, where $u,=(u_1,u_2)$, 
denotes two independent variables.
From any $2$ of these equations, eq.(\ref{inveb}), the upper 
integration bounds $x_1,x_2$ 
of the system of differential equations eq.(\ref{Umkehr})
can be expressed as single valued 
functions of the arbitrary arguments $u_1,u_2$.
For instance,
\begin{equation}
x_1=a_1+\frac{1}{A_1 (x_2-a_1) } \frac{\theta^2(u|u^{a_1,a})}{\theta^2(u)}
\label{inve1}
\end{equation}
and 
\begin{eqnarray}
x_2&=&-\;\frac{\Bigl[(a_2-a_1)(a_2+a_1)+\frac{1}{A_1}\frac{\theta^2(u|u^{a_1,a})}{\theta^2(u)}-\frac{1}{A_2}\frac{\theta^2(u|u^{a_2,a})}{\theta^2(u)}\Bigr]}
{2 (a_1-a_2)} \nonumber \\
&\pm&\frac{\sqrt{\Bigl[(a_2-a_1)(a_2+a_1)+
\frac{1}{A_1}\frac{\theta^2(u|u^{a_1,a})}{\theta^2(u)}-\frac{1}{A_2}\frac{\theta^2(u|u^{a_2,a})}{\theta^2(u)}\Bigr]^2-
4(a_1-a_2)\eta}}{2(a_1-a_2)} \nonumber \\
\label{inve2}
\end{eqnarray}
where 
\begin{equation}
\eta:=a_2\;a_1(a_1-a_2)-\frac{a_2}{A_1}\frac{\theta^2(u|u^{a_1,a})}{\theta^2(u)}+\frac{a_1}{A_2}\frac{\theta^2(u|u^{a_2,a})}{\theta^2(u)}
\end{equation}
Also, $A_i=\pm \frac{1}{\sqrt{e^{\pi i P P^{\prime}}f^{\prime}(a_i)}}$.

The solution can be reexpressed in terms of generalized Weierstra$\ss$ functions:
\begin{equation}
x_k^{(1,2)}=\frac{\wp_{2,2}(u)\pm \sqrt{\wp^2_{2,2}(u)+4\wp_{2,1}(u)}}{2},
\;\;k=1,2
\end{equation}
where
\begin{equation}
\wp_{2,2}(u)=\frac{(a_1-a_2)(a_2+a_1)-\frac{1}{A_1}\frac{\theta^2(u|u^{a_1,a})}{\theta^2(u)}+\frac{1}{A_2}\frac{\theta^2(u|u^{a_2,a})}{\theta^2(u)}}{a_1-a_2}
\end{equation}
and 
\begin{equation}
\wp_{2,1}(u)=\frac{-a_1a_2(a_1-a_2)-\frac{a_1}{A_2}\frac{\theta^2(u|u^{a_2,a})}{\theta^2(u)}+\frac{a_2}{A_1}\frac{\theta^2(u|u^{a_1,a})}{\theta^2(u)}}{a_1-a_2}
\end{equation}
Thus, $x_1,x_2$, that solve Jacobi's inversion problem Eq.(\ref{Umkehr}), 
are solutions of a quadratic equation \cite{Jacobi,BAKER}
\begin{equation}
U x^2-U^{\prime} x+U^{\prime\prime}=0
\end{equation}
where $U,U^{\prime},U^{\prime\prime}$ are functions of $u_1,u_2$.
In the particular case that the coefficient of $x^5$ in the quintic 
polynomial is equal to 4, $U=1,U^{\prime}=\wp_{2,2}(u),U^{\prime\prime}=
\wp_{2,1}(u)$.

The matrix elements $h_{ij},\Omega_{ij}$ can be explicitly written in
terms of the half-periods $U_r^{x,a}$. For clarity,  $U_2^{e_4,e_3}=
\int_{e_3}^{e_4}xdx/y,U_1^{e_4,e_3}=\int_{e_3}^{e_4}dx/y$ etc. The roots have been arranged 
in ascending order of magnitude and are denoted by $e_{2g},e_{2g-1},\cdots,e_0,
g=2$, so that $e_{2i},e_{2i-1}$ are respectively, $c_{g-i+1}, 
a_{g-i+1}$ and $e_0$ is $c$.
For instance, the matrix element $h_{11}=\frac{U_2^{e_4,e_3}}{2(
U_1^{e_4,e_3}U_2^{e_2,e_1}-U_1^{e_2,e_1}U_2^{e_4,e_3})}\times \pi i$, while 
$\Omega_{11}=\frac{U_1^{e_1,e_0}U_2^{e_4,e_3}-U_2^{e_1,e_0}U_1^{e_4,e_3}}{
U_2^{e_2,e_1}U_1^{e_4,e_3}-U_1^{e_2,e_1}U_2^{e_4,e_3}}$.

\subsection{A particular inversion problem of Jacobi}

An indefinite elliptic integral of the third kind can be regarded 
as a special form of a hyperelliptic integral 
\begin{equation}
\int_0^x\frac{(\alpha+\beta x)dx}{\sqrt{x(1-x)(1-\kappa^2 x)(1-\lambda^2 x)
(1-\mu^2 x)}}
\end{equation}
in the special case when two of the moduli are equal, e.g. $\lambda=\mu$, and 
therefore one can consider the following Jacobi's inversion problem
\begin{eqnarray}
u&=&\int_0^{x_1}\frac{(\alpha+\beta x)dx}{(1-\lambda^2 x)
\sqrt{x(1-x)(1-\kappa^2 x)}}+
\int_0^{x_2}\frac{(\alpha+\beta x)dx}{(1-\lambda^2 x)
\sqrt{x(1-x)(1-\kappa^2 x)}} \nonumber \\
v&=&\int_0^{x_1}\frac{(\alpha^{\prime}+\beta^{\prime} x)dx}{(1-\lambda^2 x)
\sqrt{x(1-x)(1-\kappa^2 x)}}+
\int_0^{x_2}\frac{(\alpha^{\prime}+\beta^{\prime} x)dx}{(1-\lambda^2 x)
\sqrt{x(1-x)(1-\kappa^2 x)}} \nonumber \\
\end{eqnarray}
For a convenient choice of the constants $\alpha,\beta,\alpha^{\prime},
\beta^{\prime}$ the solution of the above Jacobi's inversion problem 
can be expressed in terms of genus 1, Jacobi theta functions
\begin{eqnarray}
\pm \kappa \sqrt{x_1 x_2}&=&\frac{\theta(a)e^{-v}\theta(u-a)-e^v \theta(u+a)}
{\theta_1(a)e^{-v}\theta_1(u-a)+e^v\theta_1(u+a)} \nonumber \\
\frac{\kappa}{\kappa^{\prime}}\sqrt{(1-x_1) (1-x_2)}&=&
\frac{\theta_3(a)e^{-v}\theta_2(u-a)-e^v\theta_2(u+a)}
{\theta_1(a)e^{-v}\theta_1(u-a)+e^v\theta_1(u+a)} \nonumber \\
\frac{1}{\kappa^{\prime}}\sqrt{(1-\kappa^2 x_1)(1-\kappa^2 x_2)}
&=&\frac{\theta_2(a)e^{-v}\theta_3(u-a)-e^v\theta_3(u+a)}
{\theta_1(a)e^{-v}\theta_1(u-a)+e^v\theta_1(u+a)} \nonumber \\
\end{eqnarray}

\section{Transformation theory of elliptic functions and modular equations}
\label{MODULAR}

One of the applications supplied by the transformation theory of Elliptic 
functions, 
which is of great importance in Number theory, are the modular equations 
described below \cite{NHA,NHA1,CGJAC}. 

For a rational solution of the differential equation
\begin{equation}
\frac{dy}{\sqrt{(1-y^2)(1-e_1^2y^2)}}=C\frac{dx}{\sqrt{(1-x^2)(1-e^2 x^2)}}
\label{TRANMOD}
\end{equation}
the necessary conditions among the periods 
\begin{eqnarray}
K(e_1)&=& a_0 C K(e)+a_1 C i K^{\prime}(e) \nonumber \\
iK^{\prime}(e_1)&=& b_0 C K(e)+ b_1 C i K^{\prime}(e) \nonumber \\
\end{eqnarray}
with the period ratios (moduli) of the associated modular theta functions 
being given by
\begin{equation}
\tau=\frac{b_0+b_1 \tau^{\prime}}{a_0+a_1\tau^{\prime}}
\end{equation}
are also sufficient,
when 
\begin{equation}
a_0 b_1-a_1 b_0=n
\label{modcorre}
\end{equation}
is a positive integer number.
The integer $n$ is called the degree of 
transformation.

Equation (\ref{modcorre}) for $a_0,b_1,a_1,b_0\in Z$ when 
viewed as the determinant of a matrix $\in GL(2,Z)$, sometimes is called a {\em modular correspondence of
level n}. 

It  can be shown that the {\em inequivalent reduced forms of modular 
correspondences} $$\left( \begin{array}{cc}
a_0 & a_1 \\ 
b_0 & b_1%
\end{array}%
\right),$$ are of the form 
  $$\left( \begin{array}{cc}
q & 0 \\ 
16\xi & q^{\prime}%
\end{array}%
\right)$$
where $q$ a positive part of $n$ represents, $q^{\prime}:=\frac{n}{q}$, 
and $0 \leq \xi \leq q^{\prime}-1$. For instance for $n=p$ a prime number, 
there are $p+1$ inequivalent reduced forms of the form \footnote{In 
a more familiar notation these classes of inequivalent reduced forms 
are $$\alpha_i=\left( \begin{array}{cc}
a & b \\ 
0 & d%
\end{array}%
\right), ad=n,\; (a,b,d)=1,\;0\leq b <d\; {\rm and}\; a,b,d\in Z$$.}
$$\left( \begin{array}{cc}
1 & 0 \\ 
0 & p%
\end{array}%
\right),\left( \begin{array}{cc}
1 & 0 \\ 
16 & p%
\end{array}%
\right), 
\left( \begin{array}{cc}
1 & 0 \\ 
16.2 & p%
\end{array}%
\right),
\cdots \left( \begin{array}{cc}
1 & 0 \\ 
16(p-1) & p%
\end{array}%
\right),
\left( \begin{array}{cc}
p & 0 \\ 
0 & 1
\end{array}
\right)$$

Also the multiplication factor $C$ in Eq.(\ref{TRANMOD}) is given by
\begin{equation}
C=\frac{1}{q}\frac{K(e_1)}{K(e)}
\label{MULTI}
\end{equation}
which  for a degree of transformation that is a prime number ($n=p$) is equal 
to $\frac{K(e_1)}{K(e)}$ or $(1/n)\frac{K(e_1)}{K(e)}$.
Equation (\ref{MULTI}) can be re-expressed in terms of Jacobi theta functions as  follows:
\begin{equation}
C=\frac{1}{q}\frac{\vartheta_3^2(0,\tau)}{\vartheta_3^2(0,\frac
{q\tau-16\xi}{q^{\prime}})}
\end{equation}

The modular equations are equations relating the Jacobi modulus $e(p\tau)$ to 
$e(\tau)$ which are of the form 
\begin{equation}
F_p\left[\left(\frac{2}{p}\right)\sqrt[4]{e(\tau)},\sqrt[4]{e\left(\frac{\tau-16\xi}{p}\right)}\right]=0
\end{equation}
where $\left(\frac{2}{p}\right)$ denotes the Legendre symbol \footnote{
$\left(\frac{2}{a_0}\right)=e^{\frac{a_0^2-1}{8}i\pi}$.} .
Equivalently, modular equations can be written in terms of 
the absolute modular invariant function $j(\tau)$, and relate the reduced 
absolute 
modular invariant $j^{*}$ to $j$
by polynomial equations of the form
\begin{equation}
\Phi_p(j^{*},j)=0
\end{equation}
where $j^{*}:=j. \alpha_i=j\left(\frac{a\tau+b}{d}\right)$
The explicit form of $\Phi_2(j^{*},j)=0$, has been given in \cite{Yui}.

\section{Conditions for radii for spherical null geodesics with a cosmological 
constant and differential equations of Appell's function}
\label{Appell1}

The conditions from the vanishing of the polynomial $R$ and its first derivativeresult in the equations which generalize (\ref{CON})
\begin{equation}
\Phi=\frac{r^3 \Lambda^2 a^5+Y a^3+X a\pm\sqrt{3}\sqrt{f_1}}{r^3\Lambda^2 a^4+
(2 \Lambda^2 r^5 + 6 \frac{GM}{c^2} \Lambda r^2-9 r+9 \frac{GM}{c^2})a^2+
r^4\Lambda(\Lambda r^3-3 r+6 \frac{GM}{c^2})}
\end{equation}
where
\begin{eqnarray}
f_1 &:=& r^2  ( Z a^6+r(3 \frac{GM}{c^2} K+r(2 \Lambda^3r^6-15 \Lambda^2 r^4+
9\Lambda r^2+54)) \nonumber \\
&+& r^2 (54 (\Lambda r^2+2)(\frac{GM}{c^2})^2+12 r K_2 \frac{GM}{c^2}+
r^2 K_1)a^2 \nonumber \\
&+&3 (3\frac{GM}{c^2}-r)r^6 \Lambda(\Lambda r^3-3 r+6 \frac{GM}{c^2}) ) \nonumber
\end{eqnarray}

Indeed, for a vanishing cosmological constant it has the correct limit 
$\Phi \rightarrow \frac{a^2+r^2}{a}$ derived in the previous subsection.

Then the parameter ${\cal Q}$ is given by the expression
\begin{equation}
{\cal Q}=\frac{\Xi^2 r^4+\Xi^2 r^2 (a^2-\Phi^2)+
\frac{2 G M}{c^2} r \Xi^2 (\Phi-a)^2+\frac{\Lambda}{3}r^2 (r^2+a^2)\Xi^2 
(\Phi-a)^2}{\Delta_r} 
\end{equation}
where 
\begin{eqnarray}
X&:=&\Lambda^2 r^7+6 \frac{GM}{c^2}\Lambda r^4-9\frac{GM}{c^2}r^2 \nonumber \\
Y&:=&2 \Lambda^2 r^5+\frac{GM}{c^2}(6\Lambda r^2+9) \nonumber \\
Z&:=&\Lambda^3 r^6-6\Lambda^2 r^4+27 \nonumber \\
K&:=&5\Lambda^2 r^4-9\Lambda r^2-36 \nonumber \\
K_1&:=&\Lambda^3 r^6-12 \Lambda^2 r^4+18\Lambda r^2+27 \nonumber \\
K_2&=&2\Lambda^2 r^4-6 \Lambda r^2-9
\end{eqnarray}

\subsubsection{Proof of Identity Eq.(\ref{Identity})}

\begin{eqnarray}
&&\int_0^1 \frac{z^{\prime}dz^{\prime}}{\sqrt{z^{\prime}(1-\frac{z^{\prime}}
{\omega})}}\frac{1}{(1-\kappa^{\prime 2}_{+}z^{\prime})}\frac{1}{\sqrt{1-\mu^{\prime 2}z^{\prime}}} \nonumber \\
&=&\int_0^1 \frac{dz^{\prime}}{\sqrt{z^{\prime}(1-\frac{z^{\prime}}{\omega})}}
\left[\frac{z^{\prime}}{1-\kappa^{\prime 2}_{+}z^{\prime}}\right]\frac{1}{\sqrt{1-
\mu^{\prime 2}z^{\prime}}} \nonumber \\
&=&\int_0^1\frac{dz^{\prime}}{\sqrt{z^{\prime}(1-\frac{z^{\prime}}{\omega})}}
\frac{-1}{\kappa^{\prime 2}_{+}}\left[1-\frac{1}{1-\kappa^{\prime 2}_{+}z^{\prime}}\right]\frac{1}{\sqrt{1-\mu^{\prime 2}z^{\prime}}} \nonumber \\
&=&-\frac{1}{\kappa^{\prime 2}_{+}}\int_0^1 \frac{d z^{\prime}}{
\sqrt{z^{\prime} (1-\frac{z^{\prime}}{\omega})}}\frac{1}{{\sqrt{1-
\mu^{\prime 2}z^{\prime}}}}+\frac{1}{\kappa^{\prime 2}_{+}}
\int_0^1 \frac{d z^{\prime}}{\sqrt{z^{\prime} (1-\frac{z^{\prime}}{\omega})}}\frac{1}{1-\kappa^{\prime 2}_{+}z^{\prime}}
\frac{1}{\sqrt{1-\mu^{\prime 2} z^{\prime}}} \nonumber \\
\end{eqnarray}

 Picard, had developed a theory for finding  
solutions of the system of differential equations that the Appell hypergeometric function obeys. More precisely he showed, by direct substitution, 
 that solutions are provided by 
definite integrals of the form \cite{Picard}
\begin{equation}
\int_g^h u^{b_1-1} (u-1)^{b_2-1} (u-x)^{\mu-1}(u-y)^{\lambda-1} du
\end{equation}
where $g,h$ denotes two of the quantities $0,1,x,y,\infty$ and we have 
the correspondence 
\begin{equation}
b_1=1+\beta+\beta^{\prime}-\gamma,b_2=\gamma-\alpha,\mu=1-\beta,
\lambda=1-\beta^{\prime}
\end{equation}

This is the generalisation of Kummer's work who found that the 
standard hypergeometric equation has 24 solutions \cite{KUMMER}. The system 
of linear differential equations of Appell's function $F_1$ is 
\begin{eqnarray}
x(1-x)(x-y)\frac{\partial^2 F_1}{\partial x^2}+[\gamma(x-y)-(\alpha+\beta+1)x^2+
(\alpha+\beta-\beta^{\prime}+1)xy+\beta^{\prime}y]\frac{\partial F_1}{\partial x}& & \nonumber \\
-\beta y(1-y)\frac{\partial F_1}{\partial y}-\alpha\beta(x-y)F_1&=&0, \nonumber \\
y(1-y)(y-x)\frac{\partial^2 F_1}{\partial y^2}+[\gamma(y-x)-(\alpha+\beta^{\prime}+1)y^2+
(\alpha+\beta^{\prime}-\beta+1)xy+\beta x]\frac{\partial F_1}{\partial y}& & \nonumber \\
-\beta^{\prime}x(1-x)\frac{\partial F_1}{\partial x}-\alpha\beta^{\prime}(y-x)F_1&=&0, \nonumber \\
(x-y)\frac{\partial^2 F_1}{\partial x \partial y}=\beta^{\prime}\frac{\partial F_1}{\partial x}-
\beta\frac{\partial F_1}{\partial y} \nonumber \\
\end{eqnarray}
For instance the following  integral is represented as follows

\begin{eqnarray}
\int_1^{\infty} u^{\beta+\beta^{\prime}-\gamma} (u-1)^{\gamma-\alpha-1}
(u-x)^{-\beta}(u-y)^{-\beta^{\prime}}du= \nonumber \\
B(1+\beta+\beta^{\prime}-\gamma,\gamma-\alpha)x^{-\beta}y^{-\beta^{\prime}}
F_1\left(1+\beta+\beta^{\prime}-\gamma,\beta,\beta^{\prime},
1+\beta+\beta^{\prime}-\alpha,\frac{1}{x},\frac{1}{y}\right) \nonumber \\
\end{eqnarray}
and $B(p,q)=\frac{\Gamma(p) \Gamma(q)}{\Gamma(p+q)}$ denotes the 
beta function.

\end{document}